\documentclass{aa}
\usepackage{natbib}
\usepackage{amsmath}
\bibpunct{(}{)}{;}{a}{}{,}
\usepackage{graphicx}

\begin{document}

\title{Assessment of a high-resolution central scheme for the
solution of the relativistic hydrodynamics equations}

\author{Arturo Lucas-Serrano
\and Jos\'e A. Font
\and Jos\'e M$^{\underline{\mbox{a}}}$. Ib\'a\~nez
\and Jos\'e M$^{\underline{\mbox{a}}}$. Mart\'{\i} }

\offprints{Arturo Lucas-Serrano, \\
\email{arturo.lucas@uv.es}}

\institute{Departamento de Astronom\'{\i}a y Astrof\'{\i}sica,
Universidad de Valencia, Edificio de Investigaci\'on, 
Dr. Moliner, 50, 46100 Burjassot (Valencia), Spain}

\date{Received date / Accepted date}

\abstract{

We assess the suitability of a recent high-resolution central scheme
developed by \cite{kurganov} for the solution of the relativistic 
hydrodynamic equations. The novelty of this approach relies on the 
absence of Riemann solvers in the solution procedure. The computations 
we present are performed in one and two spatial dimensions in Minkowski 
spacetime. Standard numerical experiments such as shock tubes and the 
relativistic flat-faced step test are performed. As an astrophysical 
application the article includes two-dimensional simulations of the 
propagation of relativistic jets using both Cartesian and cylindrical 
coordinates. The simulations reported clearly show the capabilities of 
the numerical scheme to yield satisfactory results, with an accuracy 
comparable to that obtained by the so-called {\it high-resolution
shock-capturing} schemes based upon Riemann solvers (Godunov-type
schemes), even well inside the ultrarelativistic regime. Such central 
scheme can be straightforwardly applied to hyperbolic systems of 
conservation laws for which the characteristic structure is not 
explicitly known, or in cases where the exact solution of the Riemann 
problem is prohibitively expensive to compute numerically. Finally, we 
present comparisons with results obtained using various Godunov-type 
schemes as well as with those obtained using other high-resolution 
central schemes which have recently been reported in the literature.

\keywords{Hydrodynamics -- Methods: numerical -- Relativity -- Shock Waves}

}

\authorrunning{Lucas-Serrano et al.}
%\titlerunning{A high-resolution central scheme for relativistic hydrodynamics}
\titlerunning{A central scheme for relativistic hydrodynamics}

\maketitle

%%%%%%%%%%%%%%%%%%%%%%%%%%%%%%%%%%%%%%%%%%%%%%%%%%%%%%%%%%%%%%%%%%%%%%%%%%%%%%%%%%%%%%%%%%%%%%%%%%%%%%%%%%
\section{Introduction}
%%%%%%%%%%%%%%%%%%%%%%%%%%%%%%%%%%%%%%%%%%%%%%%%%%%%%%%%%%%%%%%%%%%%%%%%%%%%%%%%%%%%%%%%%%%%%%%%%%%%%%%%%%

Relativistic (magneto) hydrodynamical flows are present in many
astrophysical scenarios involving compact objects such as neutron
stars or black holes. The production of relativistic radio jets in
active galactic nuclei, with flow velocities as large as 99\% of
the speed of light, involves an accretion disk around a rotating
black hole \citep{begelman84}. The explosive collapse of the core
of a massive star to a neutron star in type II/Ib/Ic supernovae
contains fluid moving at relativistic speeds and strong shocks
(see e.g. \citet{mueller98} and references therein). Scenarios such as
the collapse of massive stars to black holes in collapsars, or 
coalescing neutron star binaries have been proposed as possible 
candidates to power $\gamma$-ray bursts, with bulk Lorentz factors 
larger than about 100 (see e.g. \citet{piran99,aloy00} and references 
therein).

A powerful way to improve our understanding of the physical
mechanisms which operate in those astrophysical systems is through
(magneto) hydrodynamical relativistic simulations. Numerical
simulations of relativistic flows, both in Minkowski spacetime as
in strong gravitational field scenarios, have received considerable 
attention in recent years (for a comprehensive list of references the 
reader is addressed to the recent reviews of \citet{mm99} and
\citet{font2003}). A consensus that has slowly emerged is that a
class of conservative finite-difference schemes based upon Riemann
solvers is particularly well-suited to solve the hyperbolic system
of the general relativistic hydrodynamic equations. Such schemes
are commonly known as Godunov-type schemes, a class of the
so-called {\it high-resolution shock-capturing} (HRSC) schemes
\citep{toro}. The knowledge of the characteristic speeds of the
system of equations, i.e., the eigenvalues of the Jacobian
matrices associated with the fluxes of the equations, together
with, in most cases, the corresponding eigenvectors, is
the key building block of any Riemann solver. A hydrodynamics code 
must be able, in particular, to resolve complex flows in which 
strong interacting shocks could arise. HRSC schemes written in 
conservation form have proven to fulfill the important requirement 
of capturing the profiles of strong discontinuities in a few numerical 
zones without introducing spurious oscillations.

The extension of HRSC schemes based on Riemann solvers from Newtonian 
hydrodynamics to general relativity was first discussed by \citet{mim} where 
one-dimensional test simulations were performed. Further contributions 
towards completely multidimensional formulations either adopting the 3+1 
splitting of the spacetime or fully covariant formalisms followed
\citep{eulderink,banyuls,font2,papadopoulos,ibanez}. Nowadays, many of the 
astrophysical scenarios mentioned above have been investigated numerically
using the equations of relativistic hydrodynamics and state-of-the-art HRSC 
schemes (see, e.g. references in \citet{mm99} and \citet{font2003}). 
Furthermore, there is an increasing list of relativistic Riemann solvers 
available in the literature \citep{mm99}, including the exact Riemann solver 
in the special relativistic limit \citep{mm94,pons}.

The knowledge of the characteristic structure of a hyperbolic system of 
conservation laws is recommendable, irrespective of the particular algorithm 
to solve it numerically. From the theoretical point of view such knowledge 
allows to prescribe physically consistent boundary conditions and helps to 
understand the physical properties of the system. By knowing the characteristic 
structure of the relativistic hydrodynamics equations it was possible to find 
the exact solution to the Riemann problem, both one-dimensional \citep{mm94} 
and multidimensional \citep{pons}. As a result, it was possible to know how 
the tangential velocities modify the flow solution in relativity in comparison 
with the Newtonian case \citep{pons,luciano,luciano2}. 

On the other hand, an alternative approach to upwind methods for solving hyperbolic 
systems of conservation laws by means of non-oscillatory high-order symmetric 
total-variation diminishing (TVD) schemes emerged in the mid 1980s 
\citep{davis,roe,yee1,tadmor} (see also \citet{yee2} and \citet{toro} and 
references therein). Broadly speaking these approaches are based either on 
Lax-Wendroff second-order scheme with additional dissipative terms 
\citep{davis,roe,yee1} or on non-oscillatory high-order extensions of Lax-Friedrichs
first-order central scheme \citep{tadmor}. One of the nicest properties of this 
approach is that it exploits the conservation form of the Lax-Wendroff or 
Lax-Friedrichs schemes to yield the correct propagation speeds of all nonlinear 
waves appearing in the solution. Furthermore, this procedure sidesteps the use of 
Riemann solvers, which results in enhanced computational efficiency in multidimensional 
problems. Its use is, therefore, not limited to only those systems of equations where 
the characteristic information is explicitly known or to systems where the solution of 
the Riemann problem is prohibitively expensive to compute. This approach has rapidly 
developed during the last decade to reach a mature status where a number of 
straightforward central schemes of high order can be applied to any nonlinear 
hyperbolic system of conservation laws. In particular, the typical results obtained 
for the Euler equations of Newtonian hydrodynamics show a quality comparable to that 
of HRSC schemes at the expense of a small loss of sharpness of the solution at 
discontinuities \citep{toro}. An up-to-date summary of the status and applications
of this approach is discussed in \citet{toro,kurganov,tadmor-godunov}.

In recent years there have been various successful attempts at applying high-order
central schemes to solve the relativistic (magneto) hydrodynamics equations
\citep{koide0,delzanna,anninos}. In the context of special and general 
relativistic magnetohydrodynamics (MHD), Koide and coworkers \citep{koide,koide02a} 
applied a second-order central scheme with nonlinear dissipation developed by 
\cite{davis} to the study of black hole accretion and formation of relativistic jets. 
More recently \citet{delzanna} assessed a third-order convex essentially non-oscillatory 
central scheme in multidimensional special relativistic hydrodynamics, later extended 
to relativistic MHD in \cite{delzanna2}. These authors obtained results as accurate 
as those of upwind HRSC schemes in standard tests (shock tubes, shock reflection test). 
Yet another central scheme has been lately considered by \cite{anninos} in one-dimensional 
special and general relativistic hydrodynamics, where similar results to those reported 
by \cite{delzanna} are discussed. 

The aim of this paper is to assess, as a proof of principle and motivated by the recent
stir of activity on this topic, the validity of a particularly efficient finite-difference 
central scheme written in conservation form for the solution of the relativistic hydrodynamic 
equations. This scheme was developed by \cite{kurganov} and has proven very accurate for 
solving different hyperbolic systems of conservation laws, including the Newtonian hydrodynamics 
equations. To reach our aim we perform a number of one- and two-dimensional standard numerical 
experiments in flat spacetime, such as shock tubes and the relativistic version of the so-called 
flat-faced step test \citep{emery}. We also present an astrophysical application of our adopted 
central scheme, namely the propagation of a relativistic jet. The simulations show the 
remarkable capabilities of the numerical scheme to yield satisfactory results, comparable to 
those obtained by HRSC schemes based on Riemann solvers, even well inside the ultrarelativistic 
regime.

The organization of the paper is as follows: in Section~\ref{section:equations} we remind the 
reader of the form of the system of equations of special relativistic hydrodynamics. 
Section~\ref{section:scheme} describes briefly the numerical scheme we use. Next, in 
Sections~\ref{section:results1D} and~\ref{section:results2D} we present the results of our 
one- and two-dimensional simulations, respectively. Section~\ref{comparison} contains a
quantitative comparison between the central scheme and various of the most widely used
Riemann solver-based HRSC schemes available. Finally, the summary of our investigation is 
given in Section~\ref{section:summary}. In all simulations presented we choose units such 
that the speed of light $c$ is unity.

%%%%%%%%%%%%%%%%%%%%%%%%%%%%%%%%%%%%%%%%%%%%%%%%%%%%%%%%%%%%%%%%%%%%%%%%%%%%%%%%%%%%%%%%%%%%%%%%%%%%%%%%%%
\section{Relativistic Hydrodynamic Equations}
\label{section:equations}
%%%%%%%%%%%%%%%%%%%%%%%%%%%%%%%%%%%%%%%%%%%%%%%%%%%%%%%%%%%%%%%%%%%%%%%%%%%%%%%%%%%%%%%%%%%%%%%%%%%%%%%%%%

In Minkowski spacetime and Cartesian coordinates $(t,x^i)$, the local conservation laws
describing the motion of a relativistic fluid can be cast as a first-order flux-conservative 
system of the form
\begin{equation}
\frac {\partial {\bf U}({\bf w})} {\partial t} +
\frac {\partial {\bf f}^{i}({\bf w})} {\partial x^{i}} = 0.
\label{F}
\end{equation}
\noindent
(Latin indices run from 1 to 3.) In the above equations ${\bf U}$ and ${\bf f}^i$ are, 
respectively, the state vector and the flux vector along direction $x^i$ and are defined as
 \begin{eqnarray}
 {\bf U}({\bf w})  &=& (D, S^i, \tau),
 \\
 {\bf f}^{i}({\bf w}) & = &  (D v^{i},
 S^j v^{i} + p \delta^{ij}, S^i-Dv^i).
 \end{eqnarray}
Here, $\delta^{ij}$ is the Kronecker delta, $p$ is the fluid thermal pressure related to the 
rest-mass density $\rho$ and specific internal energy density $\varepsilon$ via an equation of 
state, $p=p(\rho, \varepsilon)$, and $v^i$ is the 3-velocity. The definitions of the evolved 
quantities (relativistic densities of mass, momentum and energy) in terms of the {\it primitive} 
variables ${\bf w}=(\rho,v_i,\varepsilon)$ are
\begin{eqnarray}
D &=& \rho W,
\\
S^i &=& \rho h W^2 v^i,
\\
\tau &=& \rho h W^2 - p - D,
\end{eqnarray}
\noindent
where
$h$ is the specific enthalpy, $h=1+\varepsilon+p/\rho$, and $W$ is the Lorentz 
factor satisfying $W \equiv u^0 = 1/\sqrt{1-v^2}$ with $v^2=v^iv_i$. The 3-velocity 
components are obtained from the spatial components of the 4-velocity as 
$v^i={u^i}/{ u^0}$.

%%%%%%%%%%%%%%%%%%%%%%%%%%%%%%%%%%%%%%%%%%%%%%%%%%%%%%%%%%%%%%%%%%%%%%%%%%%%%%%%%%%%%%%%%%%%%%%%%%%%%%%%%%
\section{Numerical Scheme}
\label{section:scheme}
%%%%%%%%%%%%%%%%%%%%%%%%%%%%%%%%%%%%%%%%%%%%%%%%%%%%%%%%%%%%%%%%%%%%%%%%%%%%%%%%%%%%%%%%%%%%%%%%%%%%%%%%%%

The time update of system (\ref{F}) from $t^n$ to $t^{n+1}$ is done using an algorithm written
in conservation form
\begin{eqnarray}
{\bf U}^{n+1}_i = {\bf U}^n_i + \frac{\Delta t}{\Delta x} ({\bf F}_{i-1/2} -
{\bf F}_{i+1/2}),
\label{conservation}
\end{eqnarray}
where index $i$ labels the numerical cells. The quantities ${\bf F}_{i-1/2}$ and ${\bf F}_{i+1/2}$
are the numerical fluxes at the cell interfaces $i-1/2$ and $i+1/2$, respectively. In Riemann 
solver-based HRSC schemes those numerical fluxes are computed by solving a family of local Riemann
problems. Central schemes such as Lax-Friedrichs or Richtmyer avoid this.

In \cite{kurganov}, the authors first construct a fully-discrete central scheme, by building 
an intermediate mesh of variable cell length, making use of the {\it local speed of propagation} 
at each cell interface $a_{i+1/2}$, defined by
\begin{equation}\label{a}
  a_{i+1/2}=\max\left\{\rho\left(\frac{\partial {\bf f}}{\partial {\bf U}}
    ({\bf U}_{i+1}^{L})\right),\rho\left(\frac{\partial {\bf f}}{\partial {\bf U}}
    ({\bf U}_{i}^{R})\right)\right\},
\end{equation}
where $\rho(A)=\max_{i}(|\lambda_{i}(A)|)$, $\lambda_{i}(A)$ being the eigenvalues of the Jacobian
matrix $A\equiv\partial{\bf f}/\partial{\bf U}$. In addition, superscripts $L$ and $R$ in the above 
equation stand for the reconstructed values of ${\bf U}$ at the left and right sides of the 
corresponding numerical cell ($i+1$ and $i$, respectively). These are computed from the reconstructed 
values of the vector of primitive variables defined previously. In particular we have implemented 
in our numerical code the third-order reconstruction procedures provided by both, the PPM scheme 
\citep{colella} and the PHM scheme \citep{marquina}.

In order to avoid the computation of the Jacobian matrix of system (\ref{F}), one can calculate 
the partial derivatives  of the flux vector with respect to the state variables numerically (see 
\citet{liu,jiang}). The construction of the resulting method is not simple, and it is only second 
order accurate in space. Its extension to higher spatial order is quite involved and, for the 
equations of relativistic hydrodynamics, many intermediate calculations are necessary due to
the root-finding procedure needed to recover the primitive variables from the conserved
quantities after a time update \citep{mm99}. The final numerical flux depends on $a_{i+1/2}$ and 
on the partial derivative of the flux vector with respect to $x_{i}$, which can be again calculated 
numerically (see \cite{kurganov} for details). Preliminary results in one-dimensional simple tests 
in Minkowski spacetime are far more inaccurate than the ones obtained when using approximate Riemann 
solvers.

However, this first version of the scheme admits a semi-discrete form, by setting $\Delta t 
\rightarrow 0$. All new variables constructed on the intermediate mesh of the original scheme 
vanish, resulting in a much simpler and robust scheme. Now, we can introduce time differencing 
again by applying a standard time discretization such as a Runge-Kutta one. In this way, one 
can obtain a fully-discrete, simple, robust, and characteristic-information-free central scheme. 
Hereafter, we will refer to this scheme as {\it Tadmor's scheme}, with a numerical flux function 
given by
\begin{equation}\label{fTad2D}
  {\bf F}_{i+1/2}=\frac{1}{2}\left[{\bf f}({\bf U}_{i+1}^L)+{\bf f}
({\bf U}_i^R)\right]-\frac{a_{i+\frac{1}{2}}}{2}\left[{\bf U}_{i+1}^L-{\bf U}_i^R\right].
\end{equation}
This numerical flux depends only on the local propagation speeds $a_{i+1/2}$ and, due to 
its simple form, it can be implemented and extended to any spatial order straightforwardly.
The reader is addressed to the original article of \citet{kurganov} for a
deeper description of the numerical scheme.

%%%%%%%%%%%%%%%%%%%%%%%%%%%%%%%%%%%%%%%%%%%%%%%%%%%%%%%%%%%%%%%%%%%%%%%%%%%%%%%%%%%%%%%%%%%%%%%%%%%%%%%%%%
\section{One-dimensional tests}
\label{section:results1D}
%%%%%%%%%%%%%%%%%%%%%%%%%%%%%%%%%%%%%%%%%%%%%%%%%%%%%%%%%%%%%%%%%%%%%%%%%%%%%%%%%%%%%%%%%%%%%%%%%%%%%%%%%%

%%%%%%%%%%%%%%%%%%%%%%%%%%%%%%%%%%%%%%%%%%%%%%%%%%%%%%%%%%%%%%%%%%%%%%%%%%%%%%%%%%%%%%%%%%%%%%%%%%%%%%%%%%
\subsection{Riemann problems}\label{STT1D}
%%%%%%%%%%%%%%%%%%%%%%%%%%%%%%%%%%%%%%%%%%%%%%%%%%%%%%%%%%%%%%%%%%%%%%%%%%%%%%%%%%%%%%%%%%%%%%%%%%%%%%%%%%

A shock tube problem is a particular Riemann problem in which the states at both sides of
a given interface are at rest. The thermodynamical variables of the fluid are discontinuous.
When the interface is removed the evolution results in four constant states separated by 
three elementary waves, a rarefaction, a contact discontinuity and a shock wave. The exact 
solution of the Riemann problem in relativistic hydrodynamics for vanishing tangential speeds 
can be found in \citet{mm94}. Compared to Newtonian hydrodynamics, in relativity the nonlinear
velocity addition yields a curved velocity profile for the rarefaction wave, whereas the Lorentz
contraction narrows the constant state in between the shock wave and the contact discontinuity. 
These effects become particularly strong in the ultrarelativistic regime, especially the latter. 
Shock tube experiments in relativistic hydrodynamics have been considered by many authors to 
calibrate numerical codes. An up-to-date summary of these efforts is presented in \citet{mm99}.

We present simulations of four Riemann problems: cases (a) and (c) in \citet{mm94}, with the 
formation of two shocks and two rarefaction waves, respectively, and the shock tube problems 1 
and 2 of \citet{mm96}, for which the solution consists of a rarefaction wave moving to the left 
and a shock wave moving to the right, with a contact discontinuity in between. The one-dimensional 
computational domain extends from $x=0$ to $x=1$. At $t=0$ the interface is located at $x=0.5$. 
We use an ideal fluid equation of state, $p=(\Gamma-1)\rho\varepsilon$, with $\Gamma=4/3$ for 
the first problem and $\Gamma=5/3$ for the rest.

We show results obtained using Tadmor's scheme together with a third order Runge-Kutta time 
discretization and the spatial cell reconstruction with which we obtain the best results,
alternatively, the parabolic reconstruction used in the PPM scheme (\citet{colella,mm96}), or 
the hyperbolic one used in the PHM scheme (\cite{phm}). The particular PPM parameters we
choose for the various tests are reported in Table~\ref{tabla:ppm_param} (see the original
article by \citet{colella} for information on the meaning of these parameters). For an effective 
comparison with HRSC methods based on (approximate) Riemann solvers, we perform the same tests 
with three of them (HLLE, Roe, and Marquina), implementing them together in our code in the way 
explained in \cite{aloy}. The basic algorithms of each of these schemes can e.g. be found in 
\citet{mm99}. We note that while we employ throughout the term ``Roe" we are actually referring 
to a Roe-type scheme, in the sense that we use arithmetic averaging in the numerical flux 
computation instead of Roe-averaging. Finally, in all figures presented in this section the 
solid lines indicate the exact solution and the different symbols (plus signs, crosses, and 
circles) indicate the numerical solution for the pressure, density and velocity, respectively.

\begin{center}
\begin{table}[t]
\centering
\caption{Values of the PPM reconstruction parameters used in some of the simulations
presented in Sections~\ref{section:results1D} and~\ref{section:results2D}.}
\label{tabla:ppm_param}
\begin{tabular}{cccccccc}
\hline
Test & $K_0$ & $\eta^{(1)}$ & $\eta^{(2)}$ & $\epsilon^{(1)}$ & $\omega^{(1)}$ & 
$\omega^{(2)}$ & $\epsilon^{(2)}$ \\
 \hline
Problem 2 &  1.0 & 5.0  & 0.05 & 0.1 & 0.52 & 10.0 & 0.5 \\
Problem 3 &  1.0 & 50.0 & 0.05 & 0.1 & 0.52 & 10.0 & 0.5 \\
Problem 4 &  1.0 & 50.0 & 0.05 & 0.1 & 0.52 & 10.0 & 0.5 \\
Problem 5 &  1.0 & 5.0  & 0.05 & 0.1 & 0.52 & 10.0 & 0.0001 \\
Problem 6 &  1.0 & 5.0  & 0.1  & 0.1 & 0.52 & 10.0 & 0.5 \\
Jet       &  1.0 & 1.0  & 0.05 & 0.1 & 0.52 & 5.0  & 0.1 \\
\hline
\end{tabular}
\end{table}
\end{center}

%%%%%%%%%%%%%%%%%%%%%%%%%%%%%%%%%%%%%%%%%%%%%%%%%%%%%%%%%%%%%%%%%%%%%%%%%%%%%%%%%%%%%%%%%%%%%%%%%%%%%%%%%%
\subsubsection{Problem 1}
%%%%%%%%%%%%%%%%%%%%%%%%%%%%%%%%%%%%%%%%%%%%%%%%%%%%%%%%%%%%%%%%%%%%%%%%%%%%%%%%%%%%%%%%%%%%%%%%%%%%%%%%%%

\begin{figure}[t]
  \begin{center}
  \includegraphics[angle=-90,width=0.4\textwidth]{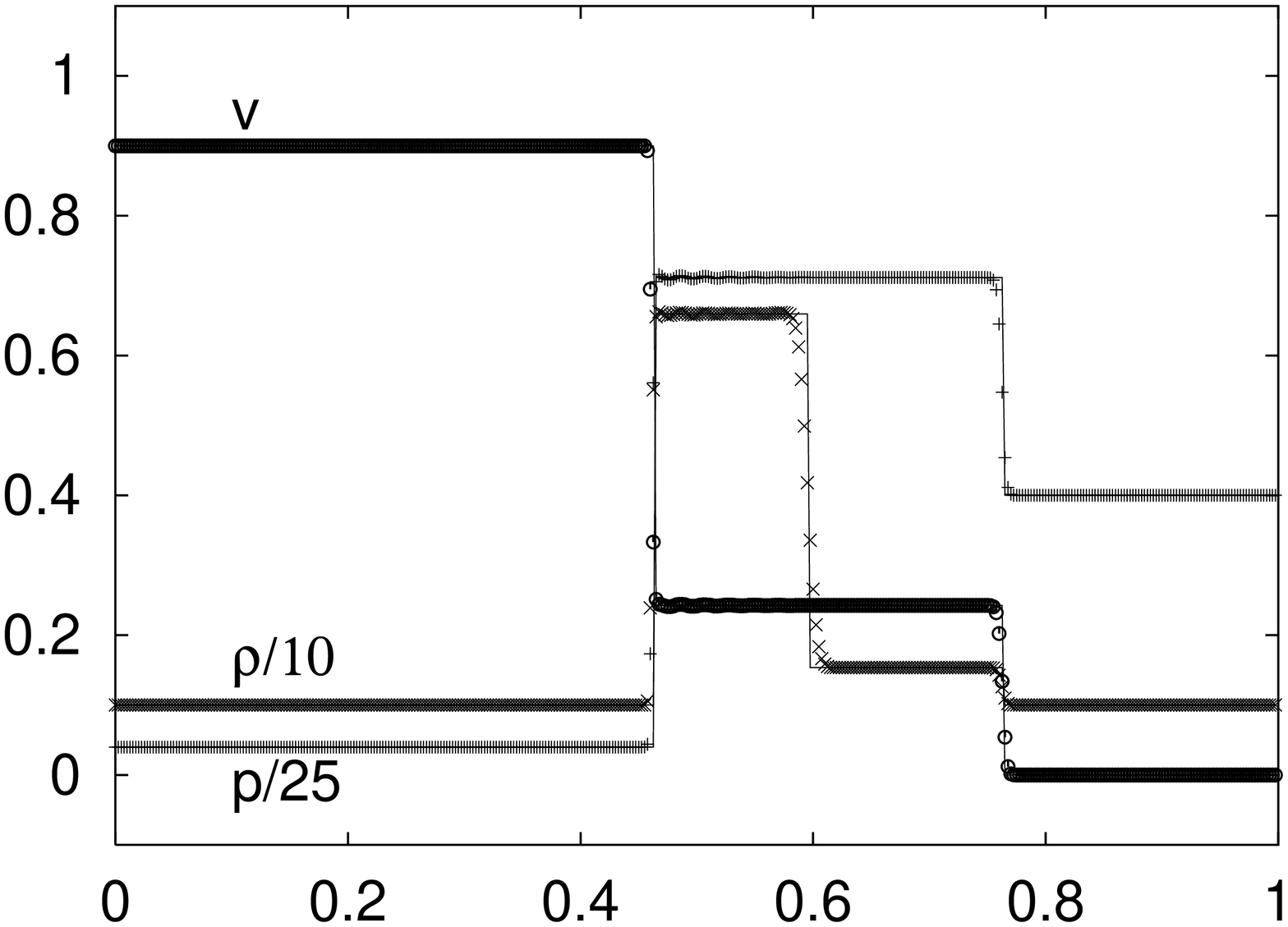}\\
  \includegraphics[angle=-90,width=0.4\textwidth]{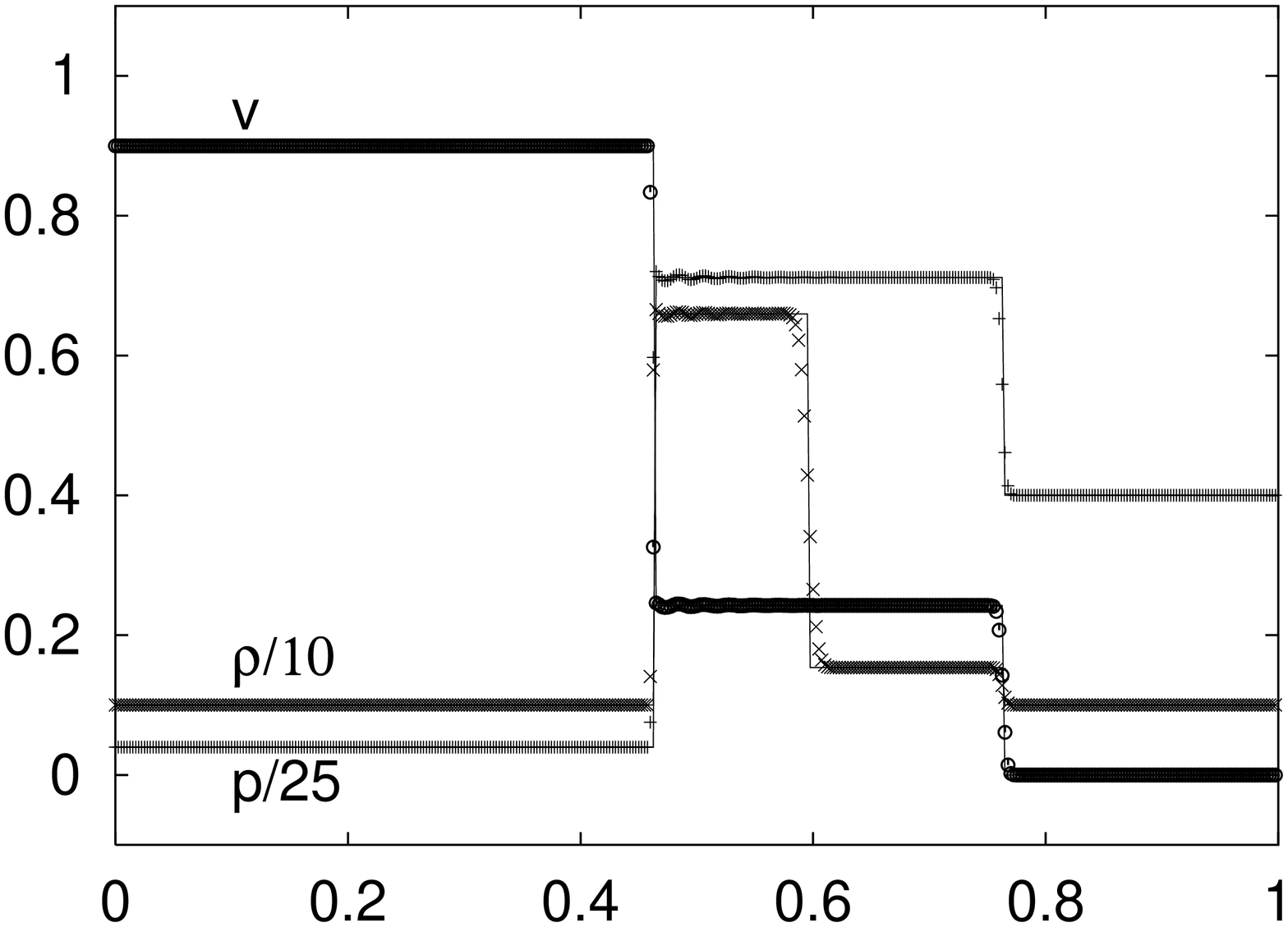}
  \caption{Tadmor's scheme (upper panel) and HLLE (lower) in the relativistic shock tube problem 1 
at $t=0.4$ using CFL=0.5 and PHM spatial reconstruction. Normalized profiles of density, pressure 
and velocity for the computed and exact (solid lines) solutions on an equally spaced grid of 400 zones.}
  \label{figure:2C}
  \end{center}
\end{figure}

The initial states are $p_{\rm L}=1.0, \rho_{\rm L}=1.0, v_{\rm L}=0.9$ (left) and $p_{\rm R}=10.0, 
\rho_{\rm R}=1.0, v_{\rm R}=0.0$ (right). Figure~\ref{figure:2C} shows the normalized profiles of the 
pressure, density and velocity at time $t=0.4$ on an equally spaced grid of 400 zones. The top 
panel corresponds to Tadmor's scheme and the bottom panel to the HLLE Riemann solver. In both cases 
we use PHM reconstruction and a CFL number of 0.5. At $t=0.4$ the left shock wave is located at 
$x=0.465$, the contact discontinuity at $x=0.6$ and the right shock is located at $x=0.76$.

The conservative central scheme we are using captures properly the location and propagation 
speeds of the different waves, with an accuracy comparable to the HLLE Riemann solver (results 
with other approximate Riemann solvers, not shown in the figure, are also similar). In particular 
it is worth mentioning the sharp resolution attained at the discontinuities, especially at the 
shocks which are resolved with the same number of points in either scheme, thanks to the use of 
the PHM third-order reconstruction. The small amplitude oscillations observed behind the left 
shock entirely disappear when lowering the value of the CFL parameter to 0.3.

%%%%%%%%%%%%%%%%%%%%%%%%%%%%%%%%%%%%%%%%%%%%%%%%%%%%%%%%%%%%%%%%%%%%%%%%%%%%%%%%%%%%%%%%%%%%%%%%%%%%%%%%%%
\subsubsection{Problem 2}
%%%%%%%%%%%%%%%%%%%%%%%%%%%%%%%%%%%%%%%%%%%%%%%%%%%%%%%%%%%%%%%%%%%%%%%%%%%%%%%%%%%%%%%%%%%%%%%%%%%%%%%%%%

\begin{figure}[t]
  \begin{center}
  \includegraphics[angle=-90,width=0.4\textwidth]{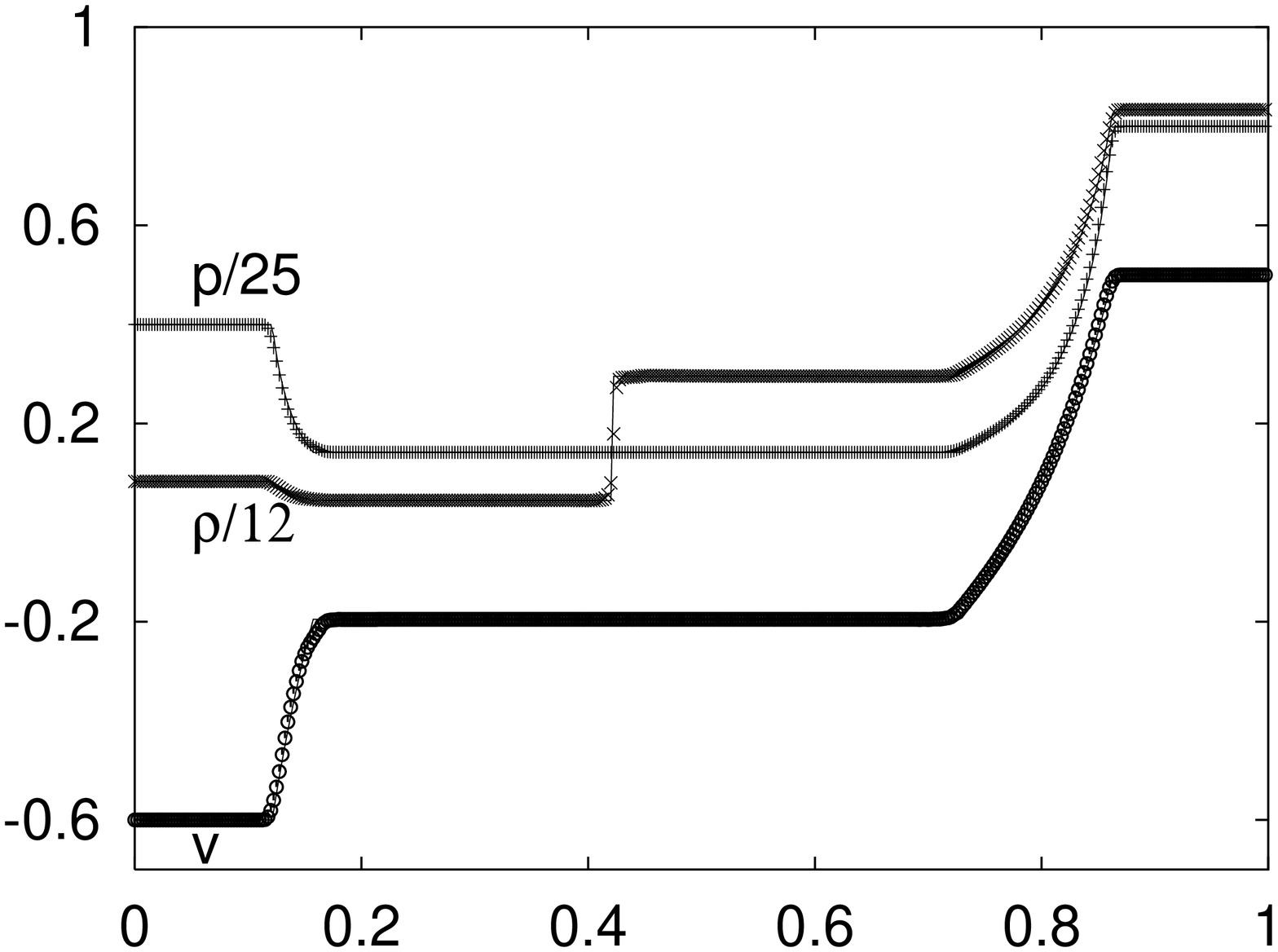}\\
  \includegraphics[angle=-90,width=0.4\textwidth]{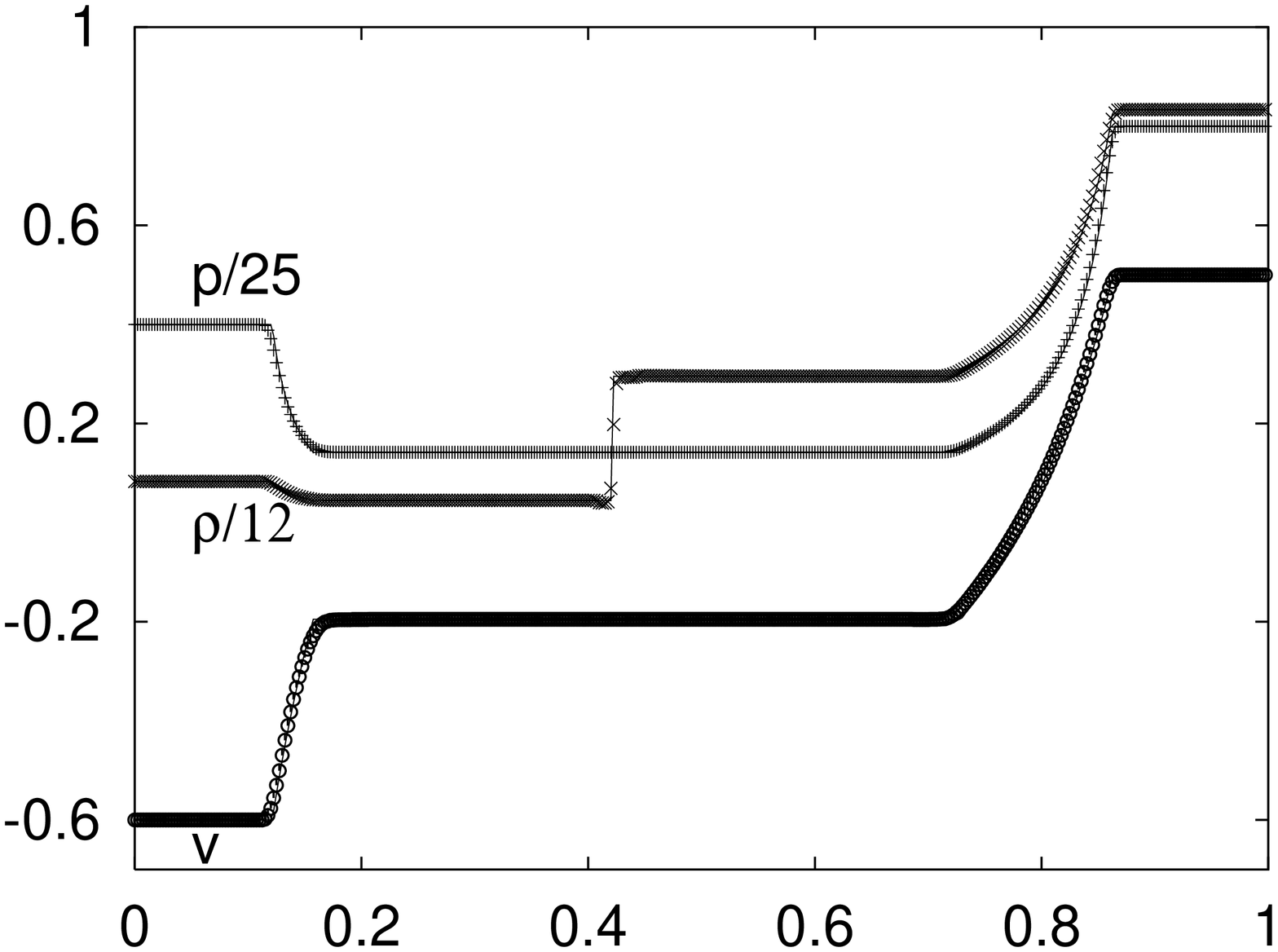}
  \caption{Tadmor's scheme (upper panel) and Roe (lower) in the relativistic shock tube 
problem 2 at $t=0.4$, using CFL=0.5 and PPM spatial reconstruction (see 
Table~\ref{tabla:ppm_param} for the specific values of the PPM parameters used). 
Normalized profiles of density, pressure and velocity for the computed and exact 
(solid lines) solutions on an equally spaced grid of 400 zones.}
  \label{figure:2R}
  \end{center}
\end{figure}

The initial states are $p_{\rm L}=10.0, \rho_{\rm L}=1.0, v_{\rm L}=-0.6$ (left) and 
$p_{\rm R}=20.0, \rho_{\rm R}=10.0, v_{\rm R}=0.5$ (right). Figure~\ref{figure:2R} shows
the normalized profiles of the pressure, density and velocity at time $t=0.4$ on an 
equally spaced grid of 400 zones. The top panel corresponds to Tadmor's scheme and 
the bottom one to Roe's approximate Riemann solver. We use PPM cell reconstruction 
with specific parameters given in Table~\ref{tabla:ppm_param}. Both schemes capture 
properly the location and propagation speeds of the different waves. It is worth pointing 
out in particular the fine performance of both methods at the contact discontinuity, a 
feature due to the use of the PPM reconstruction.

%%%%%%%%%%%%%%%%%%%%%%%%%%%%%%%%%%%%%%%%%%%%%%%%%%%%%%%%%%%%%%%%%%%%%%%%%%%%%%%%%%%%%%%%%%%%%%%%%%%%%%%%%%
\subsubsection{Problem 3}\label{section:MMI}
%%%%%%%%%%%%%%%%%%%%%%%%%%%%%%%%%%%%%%%%%%%%%%%%%%%%%%%%%%%%%%%%%%%%%%%%%%%%%%%%%%%%%%%%%%%%%%%%%%%%%%%%%%

\begin{figure}[t]
  \begin{center}
  \includegraphics[angle=-90,width=0.4\textwidth]{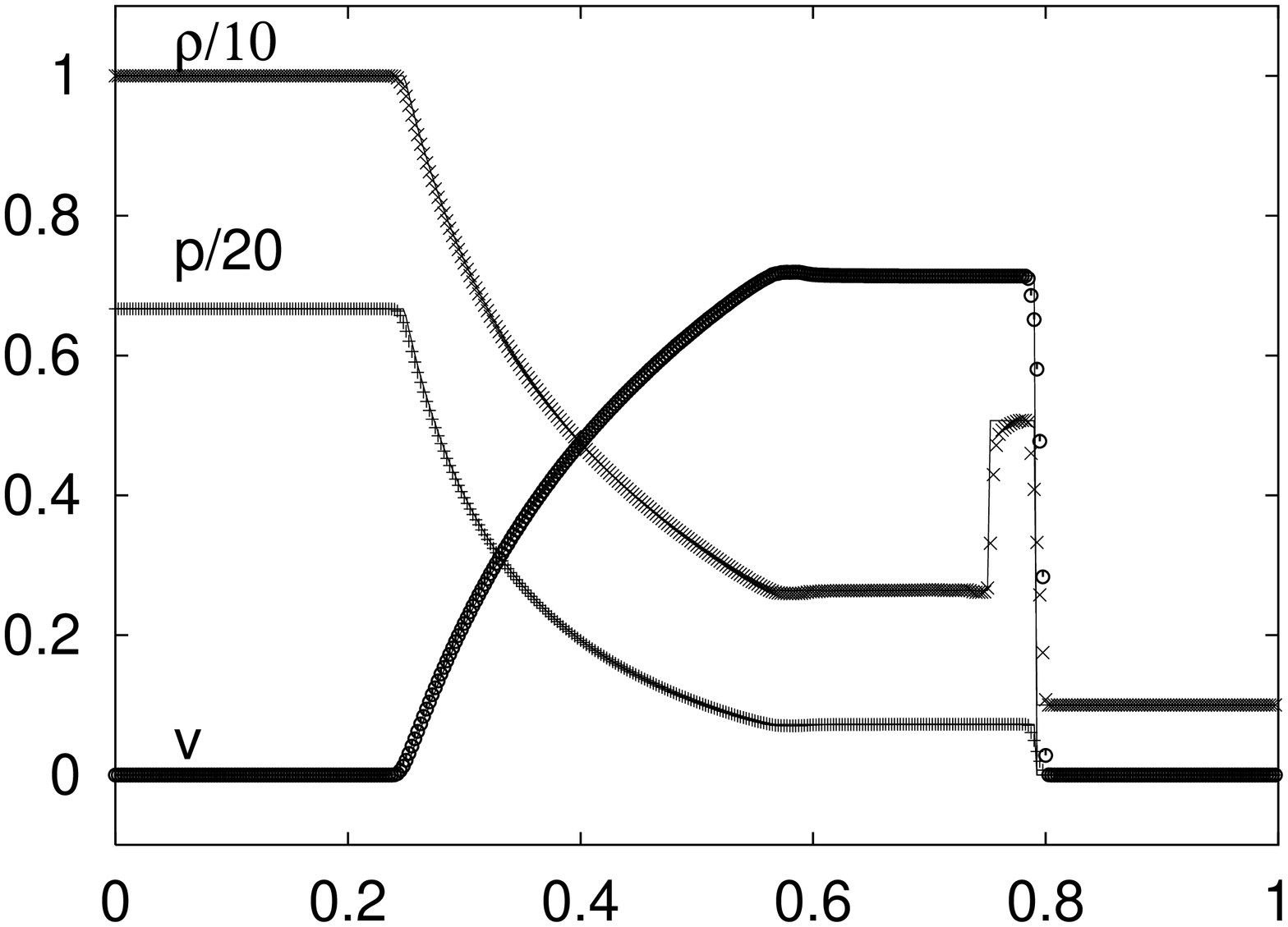}\\
  \includegraphics[angle=-90,width=0.4\textwidth]{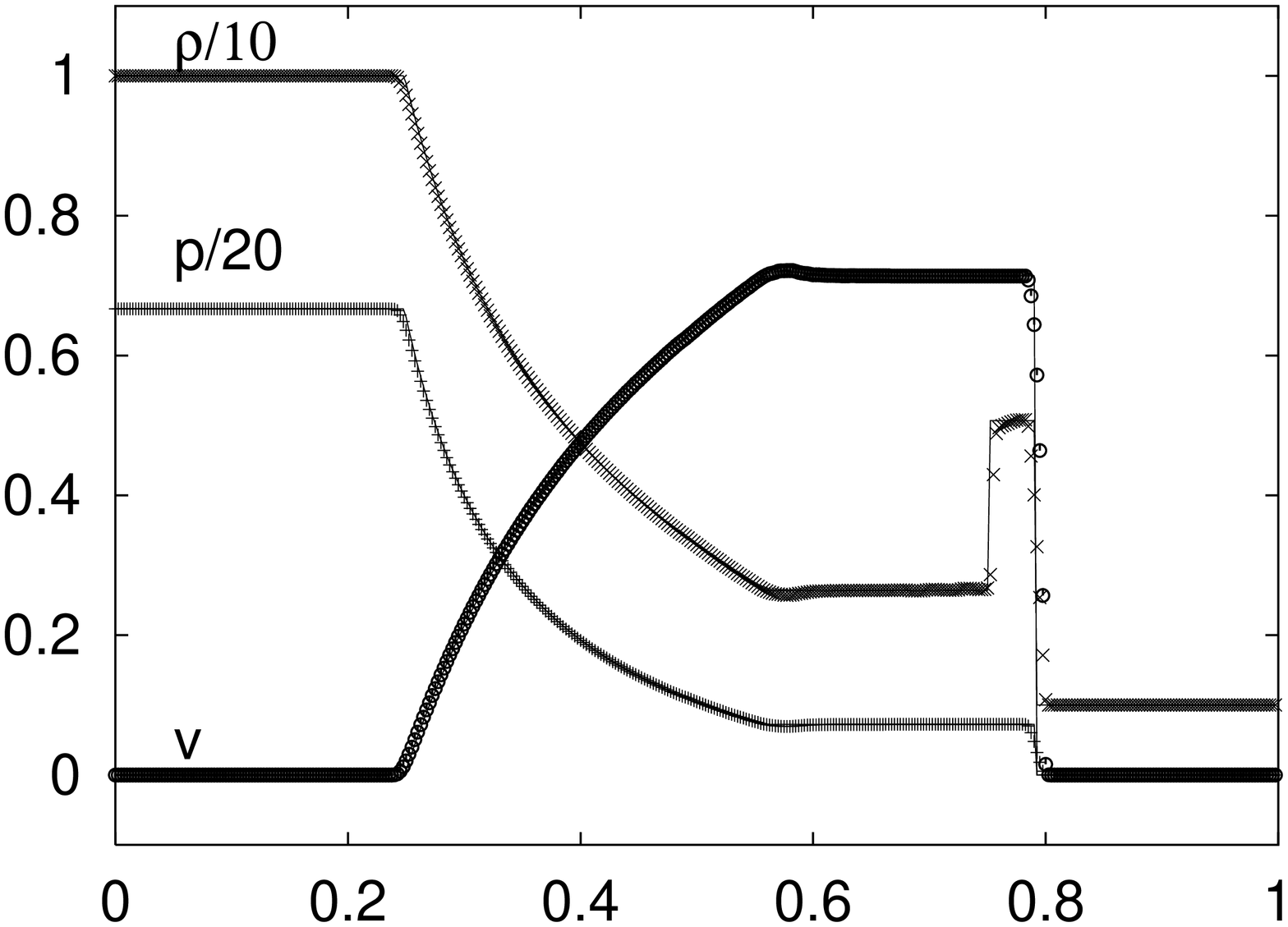}
  \caption{Tadmor's scheme (upper panel) and Marquina's flux formula (lower) in 
the relativistic shock tube problem 3 at $t=0.35$ using CFL=0.5 and PPM spatial 
reconstruction (see Table~\ref{tabla:ppm_param} for details). Normalized profiles 
of density, pressure and velocity for the computed and exact (solid lines) solutions 
on an equally spaced grid of 400 zones.}
  \label{figure:test_I}
  \end{center}
\end{figure}

The initial states are $p_{\rm L}=13.3, \rho_{\rm L}=10.0$ (left) and $p_{\rm R}=0.0, 
\rho_{\rm R}=1.0$ (right). Figure~\ref{figure:test_I} shows the normalized profiles 
of the pressure, density and velocity at time $t=0.35$ on an equally spaced grid of 
400 zones for Tadmor's scheme (top panel) and Marquina's flux formula (bottom panel). 
At $t=0.35$ the shock wave is located at $x=0.79$, the contact discontinuity at 
$x=0.75$ and the left and right corners of the rarefaction wave are located at 
$x=0.25$ and $x=0.56$, respectively. The fluid velocity behind the shock is $0.714$.
Fig.~\ref{figure:test_I} clearly shows that the location and propagation speeds of the 
different waves appearing in the solution are well captured with both schemes. Furthermore, 
the jumps in the different variables, in particular the constant state visible in the 
density in between the shock and the contact discontinuity, are also well resolved. 

The grid resolution employed in our simulations allows for a direct comparison with
the results reported in \citet{mm96}, who used a relativistic extension of the PPM 
method \citep{colella} together with the {\it exact} relativistic Riemann solver, as 
well as with those of \cite{donat98}, who employed a HRSC scheme based on Marquina's
flux formula \citep{donat96} and different high-order cell-reconstruction schemes. In 
addition, we can also compare our results with those of \cite{delzanna} and \cite{anninos} 
obtained with high-order central schemes different to the one we use. In all cases, the 
quality of our computed solution is similar to that obtained by those authors. In 
particular it is worth mentioning the sharp resolution attained at the discontinuities, 
especially at the contact discontinuity, thanks to the use of the PPM reconstruction 
with the specific parameters shown in Table~\ref{tabla:ppm_param}. As a comparison the 
third-order ENO scheme used by \citet{donat98} shows more numerical diffusion than 
Tadmor's scheme when resolving the contact discontinuity and a similar diffusion to 
the one we obtain for the case of the shock wave (see Fig. 4 in \citet{donat98}). Similarly, 
the capturing of the contact discontinuity with the central schemes of \citet{delzanna}
(see their Fig. 1) and \citet{anninos} (see their Fig. 2) appears also more diffused 
than in our case.

%%%%%%%%%%%%%%%%%%%%%%%%%%%%%%%%%%%%%%%%%%%%%%%%%%%%%%%%%%%%%%%%%%%%%%%%%%%%%%%%%%%%%%%%%%%%%%%%%%%%%%%%%%
\subsubsection{Problem 4}\label{BW}
%%%%%%%%%%%%%%%%%%%%%%%%%%%%%%%%%%%%%%%%%%%%%%%%%%%%%%%%%%%%%%%%%%%%%%%%%%%%%%%%%%%%%%%%%%%%%%%%%%%%%%%%%%

\begin{figure}[t]
  \begin{center}
  \includegraphics[angle=-90,width=0.4\textwidth]{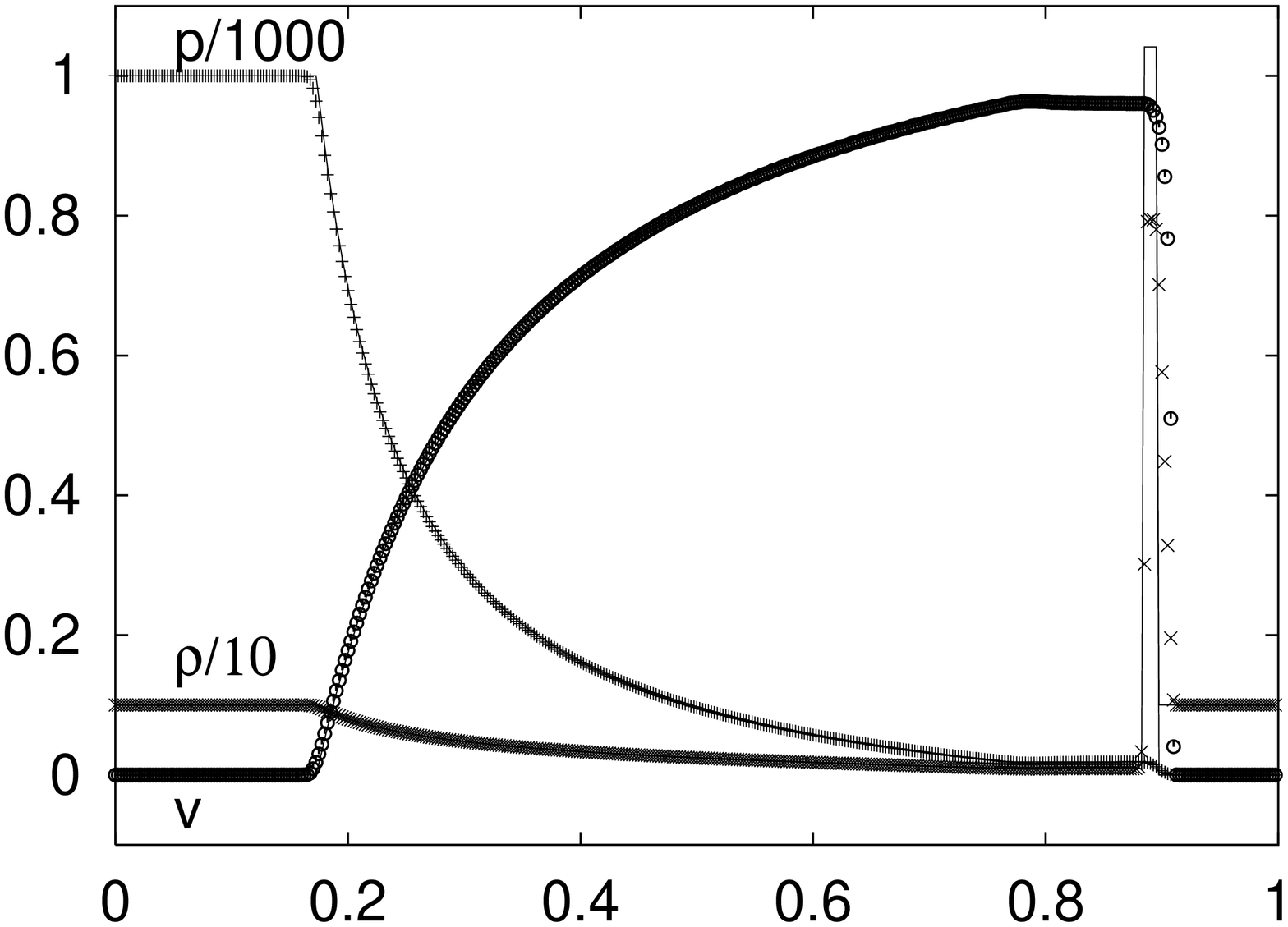}\\
  \includegraphics[angle=-90,width=0.4\textwidth]{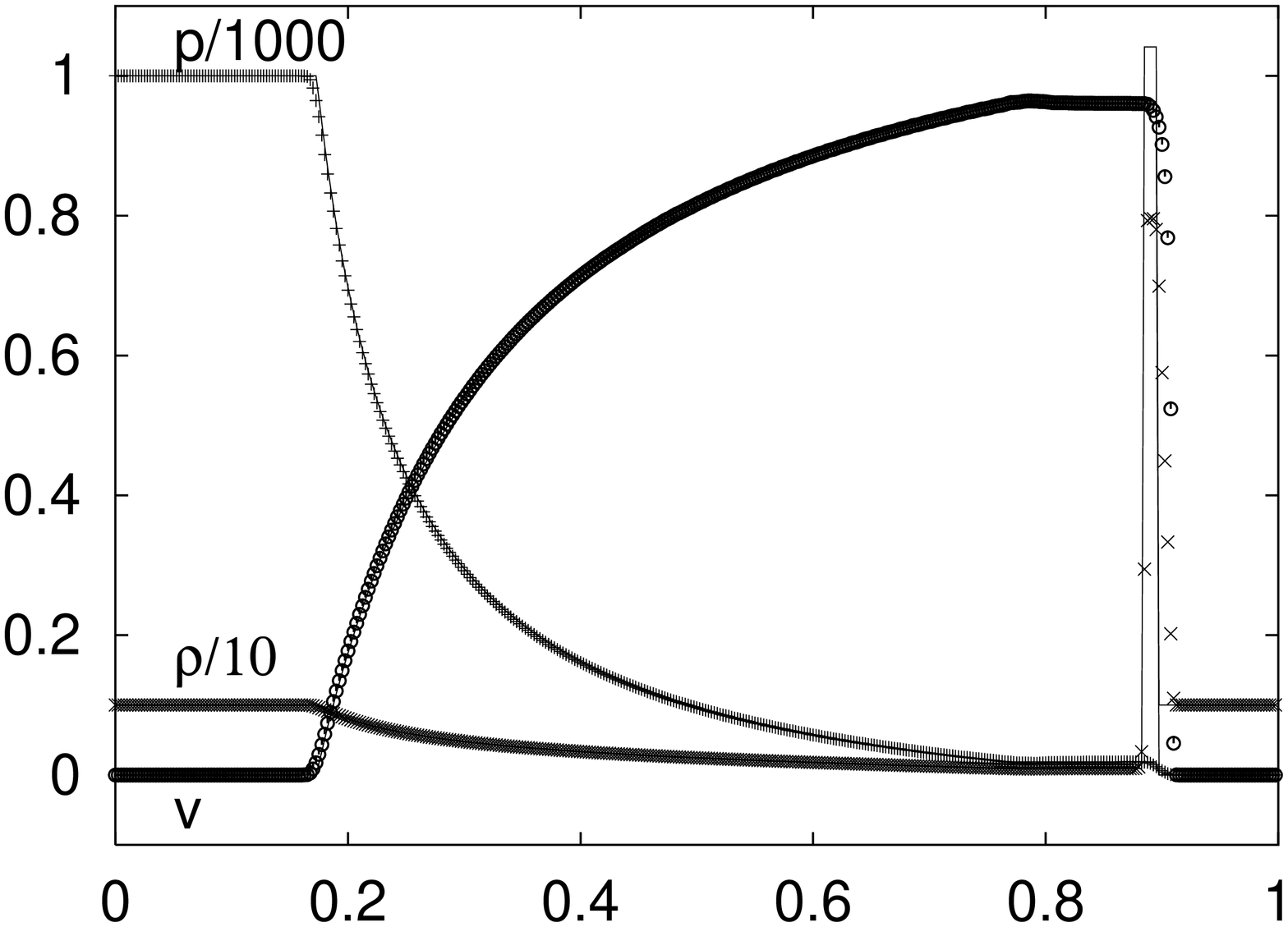}
  \caption{Tadmor's scheme (upper panel) and HLLE (lower) in the relativistic shock 
tube problem 4 at $t=0.4$ using CFL=0.4 and PPM spatial reconstruction (see 
Table~\ref{tabla:ppm_param} for details). Normalized profiles of density, 
pressure and velocity for the computed and exact (solid lines) solutions on 
an equally spaced grid of 400 zones.}
  \label{figure:BW}
  \end{center}
\end{figure}

The initial states are now $p_{\rm L}=10^3, \rho_{\rm L}=1.0$ (left) and $p_{\rm R}=10^{-2}, 
\rho_{\rm R}=1.0$ (right). Figure~\ref{figure:BW} shows the normalized profiles of the
pressure, density and velocity at time $t=0.4$ on an equidistant grid of 400 zones, 
obtained when using Tadmor's scheme (top) and HLLE (bottom) together with the PPM spatial 
reconstruction (see parameters in Table \ref{tabla:ppm_param}). The flow pattern is similar 
to that of Problem 3 but the relativistic effects make its computation much more exigent. 
The initial pressure jump of five orders of magnitude leads to the formation of a thin and 
dense shell bounded by a leading shock front and a trailing contact discontinuity - a blast 
wave. The post-shock velocity is $0.96$ ($W\sim 3.5$) while the shock speed is $0.986$ 
($W\sim 6$). Resolving the thin density plateau is a demanding test for any numerical scheme. 

As in Problem 3 we see that the central scheme we use gives the correct propagation speeds 
of the different waves, with an accuracy comparable to that of HRSC schemes based on Riemann 
solvers for the same grid resolution. By direct comparison of our results with those reported 
in \citet{mm96}, \citet{donat98}, \citet{delzanna} and \citet{anninos} we see that the overall 
agreement is similar or better. In particular, by comparing the maximum value obtained for the 
density in the thin shell, our central scheme achieves a 76\% of the exact result. The agreement 
found between a Godunov-type scheme such as HLLE and a central scheme such as Tadmor is 
remarkable. The diffusion observed in the shock wave is essentially identical to what is found 
in the previous references, while that of the contact discontinuity is now considerably smaller
due to the steepening step included in the PPM routines. With a grid of 400 zones the constant 
state in between the shock and the constant discontinuity can not yet be achieved, as in 
\citet{mm96}, \citet{donat98}, \citet{delzanna}, and \citet{anninos}. In order to get fully 
converged results with Tadmor's scheme one needs to use an equidistant grid of about 1400 zones, 
a smaller number than what is reported in \citet{donat98}. The grid requirements can obviously 
be reduced by employing adaptive mesh refinement in the region around the dense shell.

%%%%%%%%%%%%%%%%%%%%%%%%%%%%%%%%%%%%%%%%%%%%%%%%%%%%%%%%%%%%%%%%%%%%%%%%%%%%%%%%%%%%%%%%%%%%%%%%%%%%%%%%%%
\subsection{Problem 5: Shock Reflection Test}
%%%%%%%%%%%%%%%%%%%%%%%%%%%%%%%%%%%%%%%%%%%%%%%%%%%%%%%%%%%%%%%%%%%%%%%%%%%%%%%%%%%%%%%%%%%%%%%%%%%%%%%%%%

In this test an ideal cold fluid ($\varepsilon_1=0$) with velocity $v_1$ hits a wall. The fluid
is thus compressed and heats up, producing a shock which starts to propagate off the wall, leaving
the fluid behind at rest ($v_2=0$). Subscripts 1 and 2 stand for the fluid states ahead and behind
of the shock, respectively. The post-shock density is an increasing function of the initial flow
velocity. The compression ratio $\sigma\equiv\rho_2/\rho_1$ satisfies
\begin{eqnarray}
\sigma=\frac{\Gamma+1}{\Gamma-1} + \frac{\Gamma}{\Gamma-1}\varepsilon_2,
\end{eqnarray}
where $\varepsilon_2=W_1-1$. In the Newtonian limit this compression ratio is independent of the
inflow velocity. On the contrary, in the ultrarelativistic limit the density of the gas behind the
shock is unbounded ($\sigma\sim W_1$).

In our setup the computational domain covers the interval [0,1] and the wall is placed at $x=0$. 
We use a computational grid with 100 zones and an ideal fluid equation of state with $\Gamma=4/3$.
As usual in the simulations of this test the specific internal energy of the incoming fluid is set 
to a negligibly small initial value, $\varepsilon_1=10^{-10}$.

\begin{figure}[t]
  \centering
  \includegraphics[angle=-90,width=9.0cm]{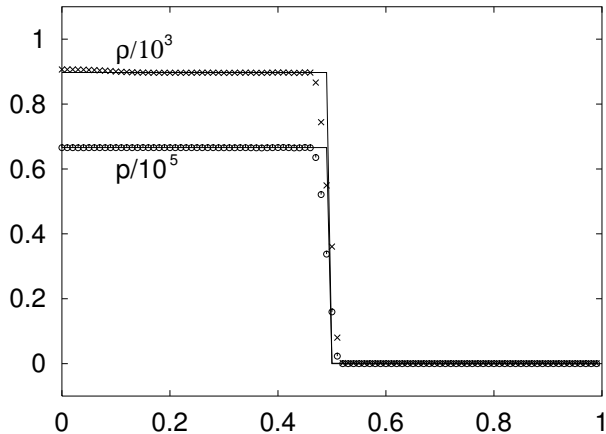}
  \caption{Tadmor's scheme in the relativistic shock reflection test using PPM reconstruction 
(see Table \ref{tabla:ppm_param} for details). Normalized profile of the pressure and the 
density for a time when the shock has propagated 0.5 units off the wall. The solid line is the 
exact solution. The symbols indicate the numerical solution obtained with a grid of 100 zones 
and CFL=0.4.}
  \label{figure:reflection}
\end{figure}

Figure~\ref{figure:reflection} shows the normalized profiles of the pressure and the density 
at a time when the shock has propagated 0.5 units off the wall. The profiles shown correspond
to an initial velocity $v_1=-0.99999$ ($W_1\sim 224$). The solid line is the exact solution 
and the crosses and circles indicate the numerical one for the pressure and the density, 
respectively. Tadmor's scheme is capable of resolving accurately the shock location and the 
associated huge jump. In comparison with results obtained with HRSC schemes based on Riemann 
solvers \citep{mm96,donat98,aloy2} or other central schemes \citep{delzanna,anninos}, the numerical 
diffusion at the shock present in our results is similar. In Fig.~8 of \citet{donat98} the shock 
is resolved in 5-6 zones, as in our central scheme, while in Fig.~2 of \citet{mm96} it is 
captured in only 3 zones. In particular, in the results of \citet{delzanna}, who only simulate 
a moderately ``cold" gas ($p=0.01$) and use 250 zones, the corner of the shock wave is not as 
sharply resolved as in our simulation (see also Fig. 9 of \citet{anninos}). In addition, the 
typical overheating error present in the density near the wall is $\sim 1\%$, somewhat lower 
than the one obtained in \citet{donat98} and \citet{delzanna}. We note once again that our 
results are obtained using the PPM reconstruction procedure (see Table \ref{tabla:ppm_param} 
for details).

\begin{figure}[t]
  \begin{center}
  \includegraphics[angle=-90,width=0.4\textwidth]{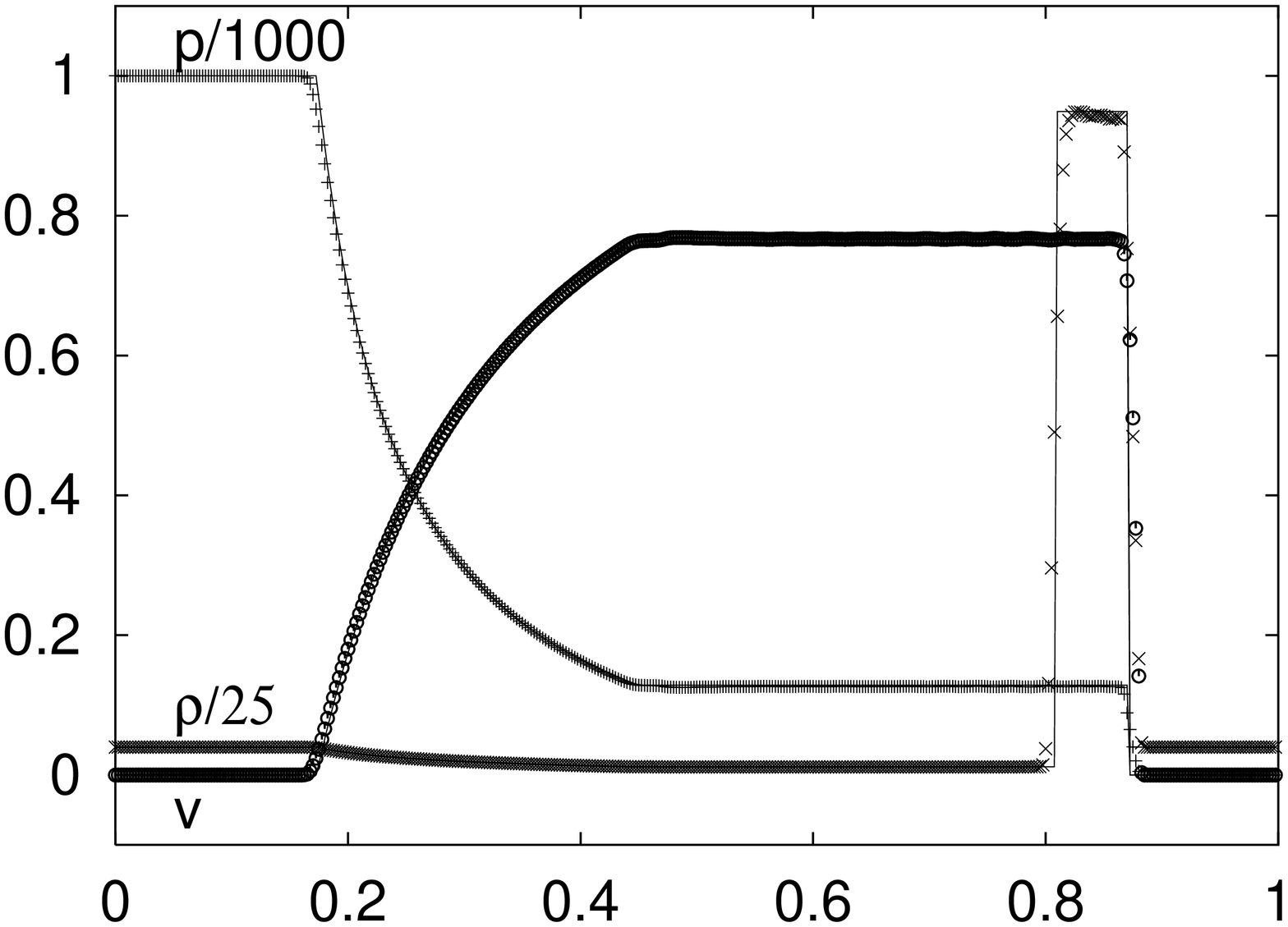}\\
  \includegraphics[angle=-90,width=0.4\textwidth]{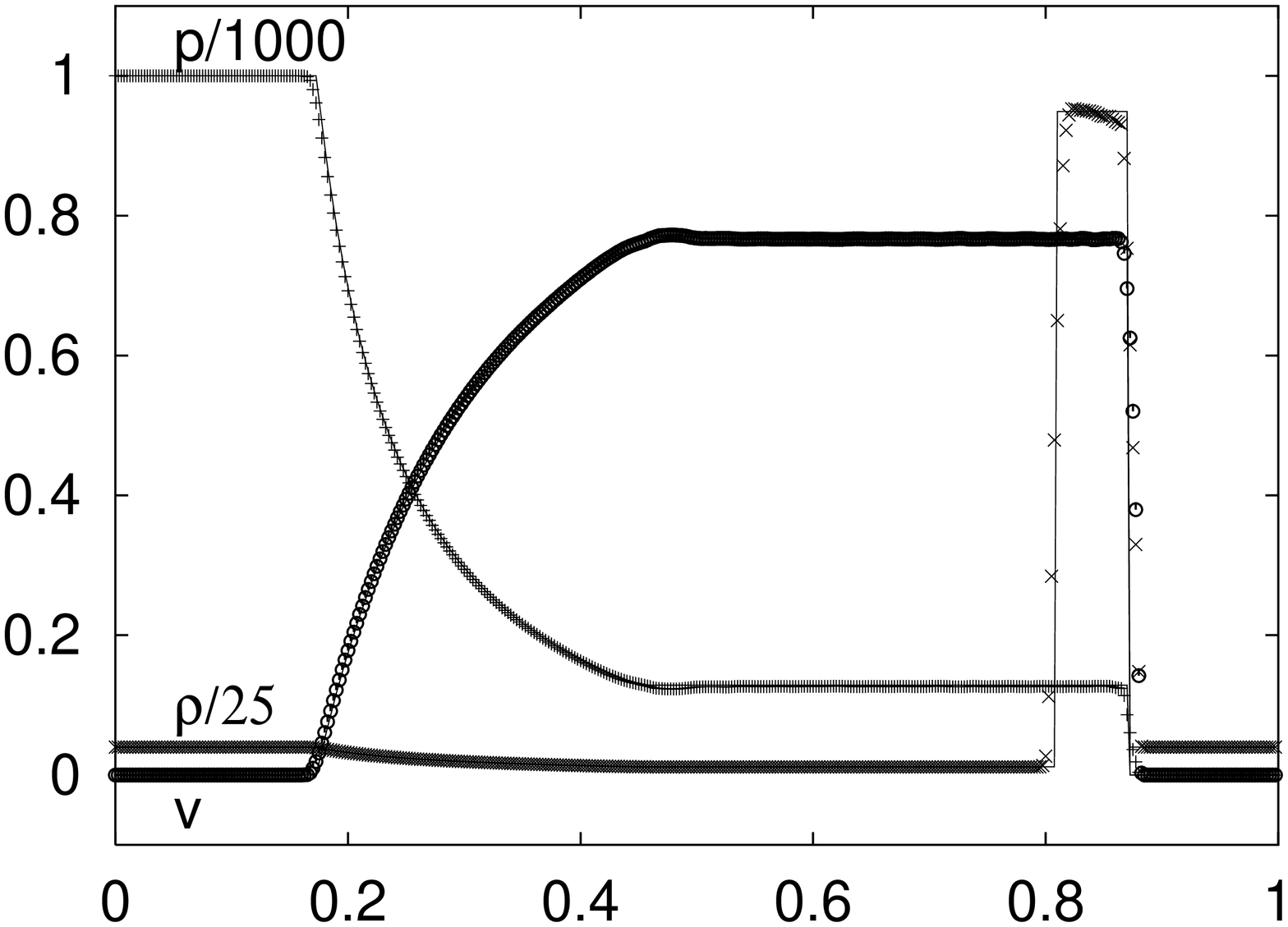}
  \caption{Tadmor's scheme (upper panel) and HLLE (lower) in the relativistic propagation 
of a blast-wave with nonzero tangential velocities at $t=0.4$ using CFL=0.5 and PPM spatial 
reconstruction (see Table \ref{tabla:ppm_param} for details). Normalized profiles of density, 
pressure and velocity for the computed and exact (solid lines) solutions on an equally spaced 
grid of 400 zones.}
  \label{figure:BWt}
  \end{center}
\end{figure}

As already shown in previous works \citep{delzanna,anninos}, high-order central schemes are 
then indeed able to achieve results comparable to Riemann solver-based schemes {\it in the 
ultrarelativistic limit} (we can reach Lorentz factors arbitrarily large, e.g. $W_1\sim
7070$, corresponding to $v=-0.99999999c$). The most important property which allows for central 
schemes such as the one presented here to succeed in the ultrarelativistic regime seems to be, 
in retrospect, the {\it conservation form} of the scheme, something which was not taken into 
account in the pioneer investigations in (ultra) relativistic hydrodynamics \citep{norman}.

%%%%%%%%%%%%%%%%%%%%%%%%%%%%%%%%%%%%%%%%%%%%%%%%%%%%%%%%%%%%%%%%%%%%%%%%%%%%%%%%%%%%%%%%%%%%%%%%%%%%%%%%%%
\subsection{Problem 6: Shock Tube Tests with non-zero tangential velocities}
%%%%%%%%%%%%%%%%%%%%%%%%%%%%%%%%%%%%%%%%%%%%%%%%%%%%%%%%%%%%%%%%%%%%%%%%%%%%%%%%%%%%%%%%%%%%%%%%%%%%%%%%%%

All shock tube problems we have considered so far involve only the velocity normal to the 
initial discontinuity. A higher level of complexity can be achieved by including the two 
other velocity components $v_y$ and $v_z$, which can be considered together as $v_\bot$, 
the velocity component defined on the perpendicular plane to the $x$-axis. As shown by
\citet{pons}, where the exact solution of such a ``multidimensional" Riemann problem was 
derived, the introduction of this new variable changes the final result of the initial 
Riemann problem.

Following the previous reference we have simulated the propagation of a relativistic blast 
wave (see~\S\ref{BW}). The final solution depends on the initial value of the tangential
velocities, which can be taken as free parameters (see \cite{pons} for details). In our 
particular case we choose $v_\bot^L=0$, $v_\bot^R=0.99$. We use the same configuration as 
in Section~\ref{STT1D} for the spatial domain and the equation of state.

In Figure~\ref{figure:BWt} we plot the results obtained using Tadmor's scheme and HLLE 
approximate Riemann solver. As in the previous tests we use a third order Runge-Kutta time 
discretization and the PPM scheme for the spatial cell-reconstruction (see 
Table \ref{tabla:ppm_param} for details on the PPM parameters). The solution plotted in 
Fig.~\ref{figure:BWt} corresponds to time $t=0.40$, and it shows the distinctive material 
shell bounded by a contact discontinuity and a shock. The shell is now thicker than in the 
case with no tangential velocities (Problem 4) and the density jump is larger. We again find 
that Tadmor's scheme captures successfully the correct wave patterns, with a very high 
accuracy at the discontinuities, due to the use of the PPM reconstruction. We note in particular
that the contact discontinuity is similarly well resolved with both schemes, HLLE and Tadmor.
It is worth emphasizing that despite the further complexity introduced by the presence of 
tangential velocities, the correct value for the density of the material shell is nevertheless 
computed very accurately.

%%%%%%%%%%%%%%%%%%%%%%%%%%%%%%%%%%%%%%%%%%%%%%%%%%%%%%%%%%%%%%%%%%%%%%%%%%%%%%%%%%%%%%%%%%%%%%%%%%%%%%%%%%
\section{2-dimensional Relativistic Hydrodynamics}
\label{section:results2D}
%%%%%%%%%%%%%%%%%%%%%%%%%%%%%%%%%%%%%%%%%%%%%%%%%%%%%%%%%%%%%%%%%%%%%%%%%%%%%%%%%%%%%%%%%%%%%%%%%%%%%%%%%%

The simplicity of Tadmor's scheme allows to implement it in multidimensions in a 
straightforward way, by using, e.g. the so-called method of lines (see e.g. \citet{toro}).
In this framework we have thus considered two different tests and an astrophysical application,
namely a two-dimensional shock tube, the relativistic version of the so-called flat-faced step 
test, and the propagation of a relativistic jet. Our results can be directly compared with 
those of previous authors \citep{donat98,marquina,delzanna}.

\begin{figure}[t]
  \centering
  \includegraphics[width=7cm]{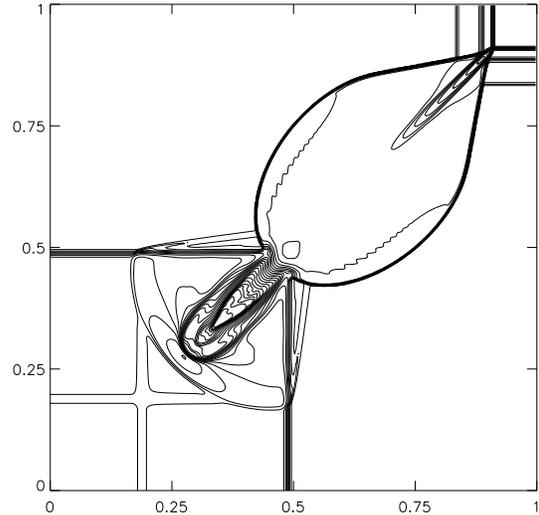}
  \caption{Tadmor's scheme with PHM reconstruction in a two-dimensional shock tube test. 
30 iso-contours of the logarithm of the density are shown. $t=0.4$, CFL=0.5 and 
$\frac{1}{\Delta x}=\frac{1}{\Delta y}=\frac{1}{400}$.}
  \label{figure:riemann2D}
\end{figure}

%%%%%%%%%%%%%%%%%%%%%%%%%%%%%%%%%%%%%%%%%%%%%%%%%%%%%%%%%%%%%%%%%%%%%%%%%%%%%%%%%%%%%%%%%%%%%%%%%%%%%%%%%%
\subsection{Two-dimensional shock tube test}
%%%%%%%%%%%%%%%%%%%%%%%%%%%%%%%%%%%%%%%%%%%%%%%%%%%%%%%%%%%%%%%%%%%%%%%%%%%%%%%%%%%%%%%%%%%%%%%%%%%%%%%%%%

In this test the computational domain is a square of side-length unity divided in four quadrants 
of equal size with constant initial states each and outflowing boundary conditions. \citet{lax-liu} 
studied the time evolution of all possible initial configurations within the framework of Newtonian 
hydrodynamics. A relativistic version of their particular configuration 12, where the four boundaries 
define two contact discontinuities and two shock waves, was proposed in \cite{delzanna}. We simulate 
the same configuration in order to compare with their results. The initial setup is

\begin{equation}
\left\{
\begin{aligned}
(\rho,v_x,v_y,p)^{NE}=&(0.1,0.0,0.0,0.01),\\
(\rho,v_x,v_y,p)^{NW}=&(0.1,0.99,0.0,1.0),\\
(\rho,v_x,v_y,p)^{SW}=&(0.5,0.0,0.0,1.0),\\
(\rho,v_x,v_y,p)^{SE}=&(0.1,0.0,0.99,1.0).\\
\end{aligned}
\right.
\end{equation}

Figure~\ref{figure:riemann2D} shows the result we obtain using Tadmor's scheme with a 
400$\times$400 grid at time $t=0.4$. We use PHM spatial reconstruction, a third 
order Runge-Kutta time discretization and a CFL number of 0.5. The solution shows two 
curved shock fronts evolving in the upper-right (NE) quadrant, and a more complicated wave 
pattern in the lower-left (SW) quadrant. The comparison with the results of \citet{delzanna}
(see their Fig. 6) reveals that all expected features in the wave solution are also 
obtained with our central scheme. The capturing of the curved shocks in the upper-right
quadrant is similarly well performed by both schemes. However, the elongated structure
appearing along the diagonal within the two shocks is more pronounced in our result (in 
closer agreement with the Newtonian results by \citet{lax-liu}). Correspondingly, concerning 
the wave structure in the lower-left quadrant we note that the location and resolution of 
the bow shock agrees in both schemes but, however, the contact discontinuity and the
structure in front of the oblique shock differ, being perhaps better resolved with
our scheme.

%%%%%%%%%%%%%%%%%%%%%%%%%%%%%%%%%%%%%%%%%%%%%%%%%%%%%%%%%%%%%%%%%%%%%%%%%%%%%%%%%%%%%%%%%%%%%%%%%%%%%%%%%%
\subsection{Relativistic flat-faced step test}
%%%%%%%%%%%%%%%%%%%%%%%%%%%%%%%%%%%%%%%%%%%%%%%%%%%%%%%%%%%%%%%%%%%%%%%%%%%%%%%%%%%%%%%%%%%%%%%%%%%%%%%%%%

A challenging test for two-dimensional hydrodynamical codes is the numerical simulation
of a supersonic flow in a wind tunnel with a flat-faced step, a test originally introduced
by \citet{emery} to compare various schemes in classical fluid dynamics. The initial 
configuration of this test is as follows: the tunnel is three units long and one unit wide. 
The step is 0.2 units high and it is located 0.6 units from the left-hand end of the tunnel. 
A Mach 3 flow (Newtonian definition) is injected through the left end of the tunnel. All 
the computational domain is initially filled with an ideal gas with $\gamma=7/5$ and constant 
density $\rho(x,y)=1.4$. The only non-vanishing initial velocity component is the horizontal 
($x$) one, which we leave as a free parameter in order to consider different regimes, from 
low Lorentz factors up to ultrarelativistic situations. Reflecting boundary conditions are 
applied along the walls of the tunnel as well as on the boundary defined by the step. 
Correspondingly, outflow (zero gradient) boundary conditions are used on the right-hand end 
of the tunnel.

\begin{figure}[t]
  \centering
  \includegraphics[width=8cm]{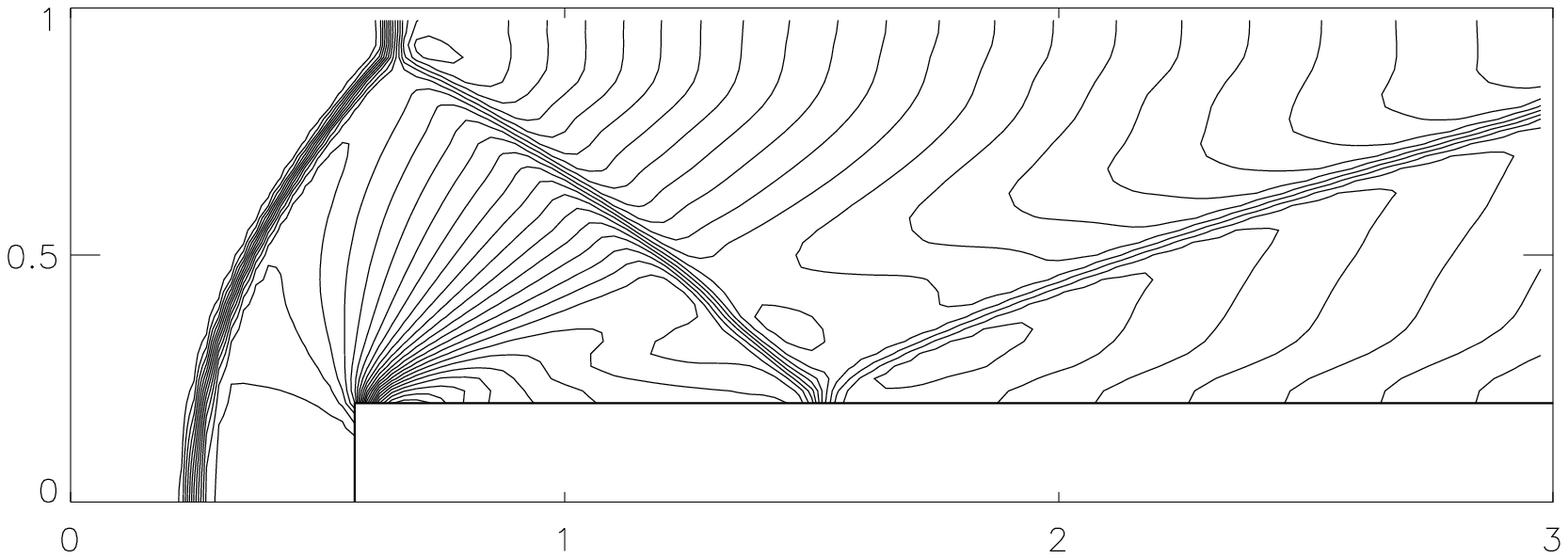}\\
  \includegraphics[width=8cm]{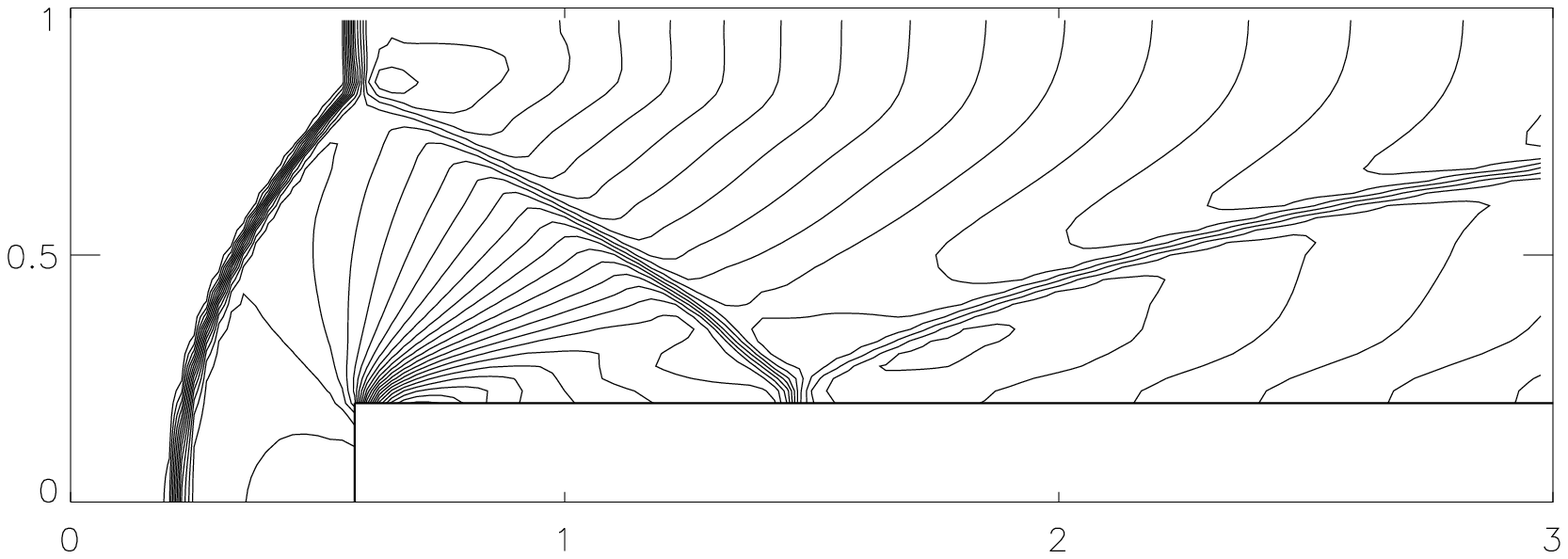}\\
  \includegraphics[width=8cm]{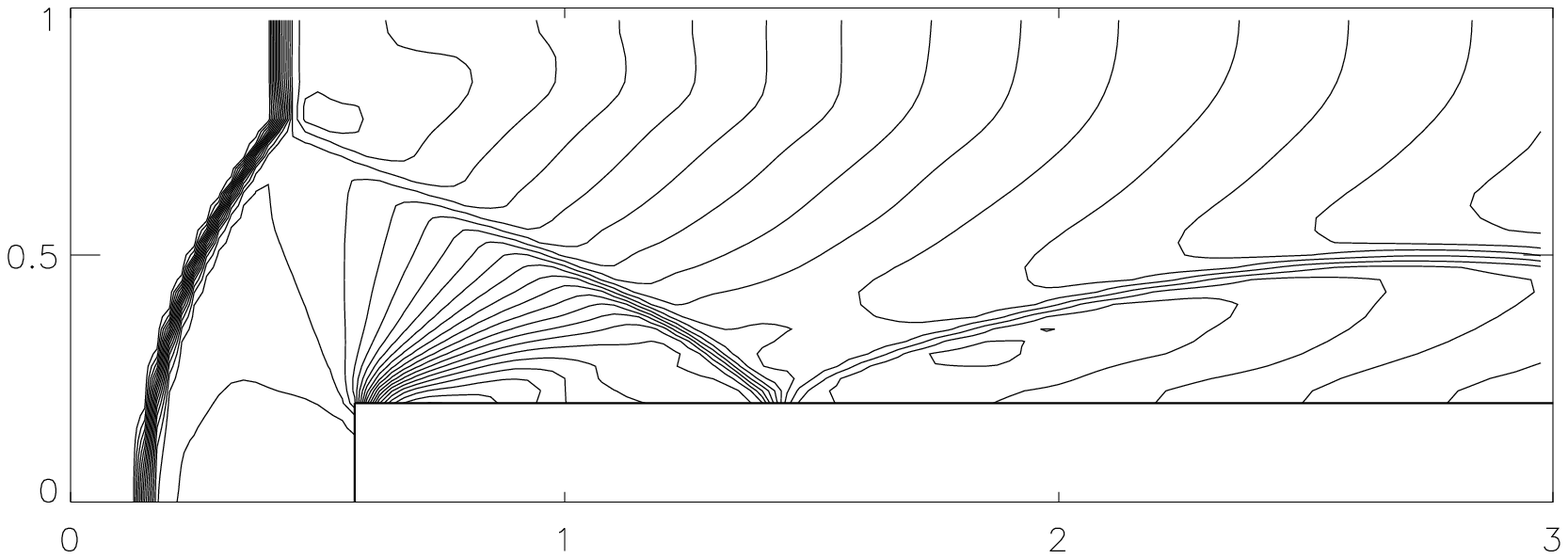}\\
  \includegraphics[width=8cm]{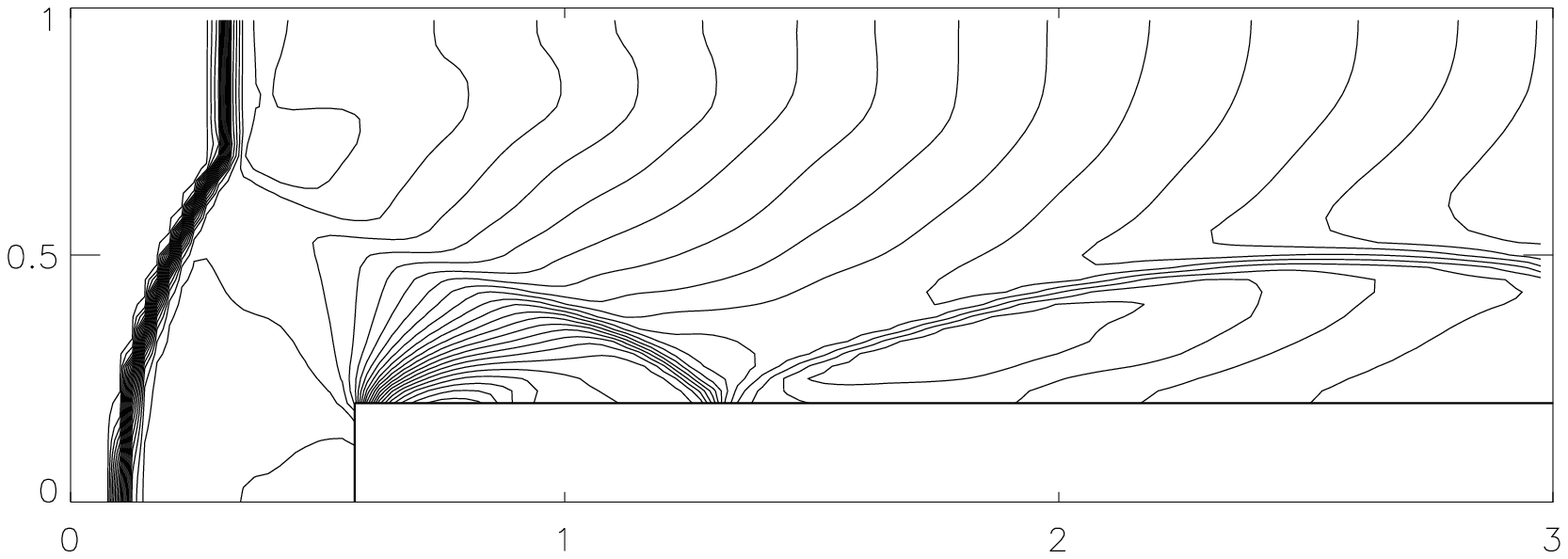}
  \caption{Tadmor's scheme with PHM reconstruction in the flat-faced step test. 
30 iso-contours of the logarithm of the density are shown. CFL=0.7 and 
$\frac{1}{\Delta x}=\frac{1}{\Delta y}=\frac{1}{40}$. From top to bottom: 
$v_x=0.1, 0.9, 0.99$, and $0.999$, and $t=55.1, 6.0, 4.45$, and $4.26$}
  \label{2Dtadmor}
\end{figure}

The corner of the step is a singular point of the flow since it is the center of a rarefaction 
fan. It is well known that linearized Riemann solvers need an entropy fix in the vicinity of 
the corner (see \citet{donat98} for details) in order to minimize the numerical errors 
generated around it which may affect the entire flow globally. We note that central schemes
such as the one we use do not need any further adjustment to pass this test without diminishing 
the quality of the results.

\begin{figure}[t]
  \centering
  \includegraphics[width=8cm]{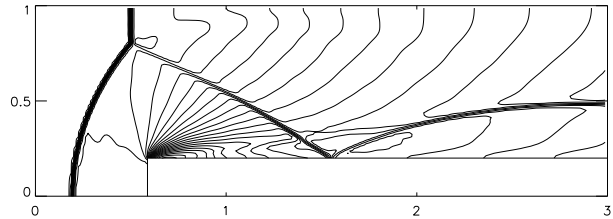}
  \caption{Tadmor's scheme with PHM reconstruction in the flat-faced step test. 
30 iso-contours of the logarithm of the density are shown. $v_x=0.99$, $t=4.5$, CFL=0.7 
and $\frac{1}{\Delta x}=\frac{1}{\Delta y}=\frac{1}{80}$}
  \label{2Dtadmor992x}
\end{figure}

Figure~\ref{2Dtadmor} plots the results obtained by Tadmor's scheme in a rectangular grid 
of 120 $x$-cells and 40 $y$-cells, using a third order Runge-Kutta time discretization and 
PHM spatial reconstruction. We note that the reconstruction is now made on the proper velocity 
components of the gas (which are unbounded), in order to avoid exceeding the speed of light 
during the recovery of the primitive quantities. From top to bottom Fig.~\ref{2Dtadmor} 
shows a snapshot in the evolution of the flow for the values of the initial velocity 
$v_x=0.1, 0.9, 0.99$, and $0.999$. The evolution of the fluid and the shock reflections pattern
are similar to the one found in the Newtonian case, but the bow shock moves faster in the 
relativistic case, eventually leaving the computational domain, the sooner the larger the 
inflow velocities are. The results obtained with Tadmor's scheme are very satisfactory, and 
again comparable to the ones obtained by Riemann solver-based methods which make use of the 
characteristic information of the system of equations (see e.g. the corresponding figures 
reported in \citet{donat98,marquina}). All special features (bow shock, rarefactions fan, etc.) 
are well resolved and correctly captured. We note that by refining the grid resolution we do 
not encounter the numerical pathologies commented in \citet{marquina}, as we show in 
Figure~\ref{2Dtadmor992x} where a grid of 240 $x$-cells and 80 $y$-cells was used. It is 
worth commenting that when using a second order spatial reconstruction such as the MC limiter 
(see \citet{toro}) and a second order Runge-Kutta time discretization, the results are still 
acceptable, but with a lower resolution at the discontinuities.

%%%%%%%%%%%%%%%%%%%%%%%%%%%%%%%%%%%%%%%%%%%%%%%%%%%%%%%%%%%%%%%%%%%%%%%%%%%
\subsection{An astrophysical application: propagation of relativistic jets}
%%%%%%%%%%%%%%%%%%%%%%%%%%%%%%%%%%%%%%%%%%%%%%%%%%%%%%%%%%%%%%%%%%%%%%%%%%%

The simulation of relativistic jets (a field of research that started about a decade ago; 
see e.g.~\citet{mmi94}) has now reached its maturity, gradually incorporating more  
elaborate ingredients such as three dimensional effects, magnetic fields, realistic equations 
of state, and emission processes (see~\citet{mm99} and references therein). As a sample of 
an astrophysical application of our central scheme, we present in this section numerical 
simulations, using both Cartesian and cylindrical coordinates, of the propagation of a 
relativistic jet in two spatial dimensions through a homogeneus environment, discussing 
briefly the main results. For the sake of comparison the initial data considered in our 
simulations are the same as those of~\citet{delzanna}. Hence, a light (jet-to-ambient 
density ratio, 0.01), relativistic (jet Lorentz factor, 7.1) jet with internal Mach number 
17.9 (highly supersonic) is injected into a homogeneous, static external medium. The jet 
and the ambient medium are in pressure equilibrium. We use an ideal gas equation of state 
with $\gamma = 5/3$ to represent both jet and ambient medium. Reflecting boundary conditions 
are used at the jet symmetry axis and a combination of outflow and reflection conditions
at the boundary above the jet inlet (at the left corner of the numerical grid). Pure 
outflow boundary conditions are imposed in the remaining boundaries. The simulations are 
performed with a resolution of 20 cells per jet radius and with a CFL of 0.3.

\begin{figure}[t]
\centering
\includegraphics[width=8cm]{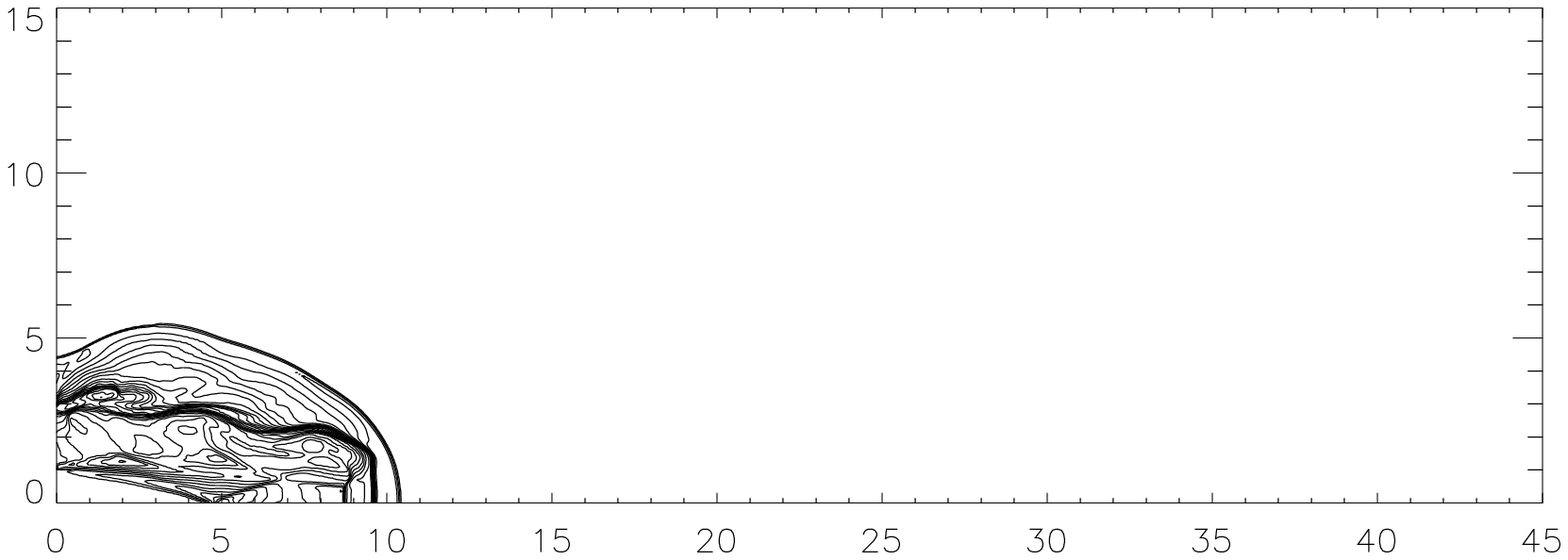}\\
\includegraphics[width=8cm]{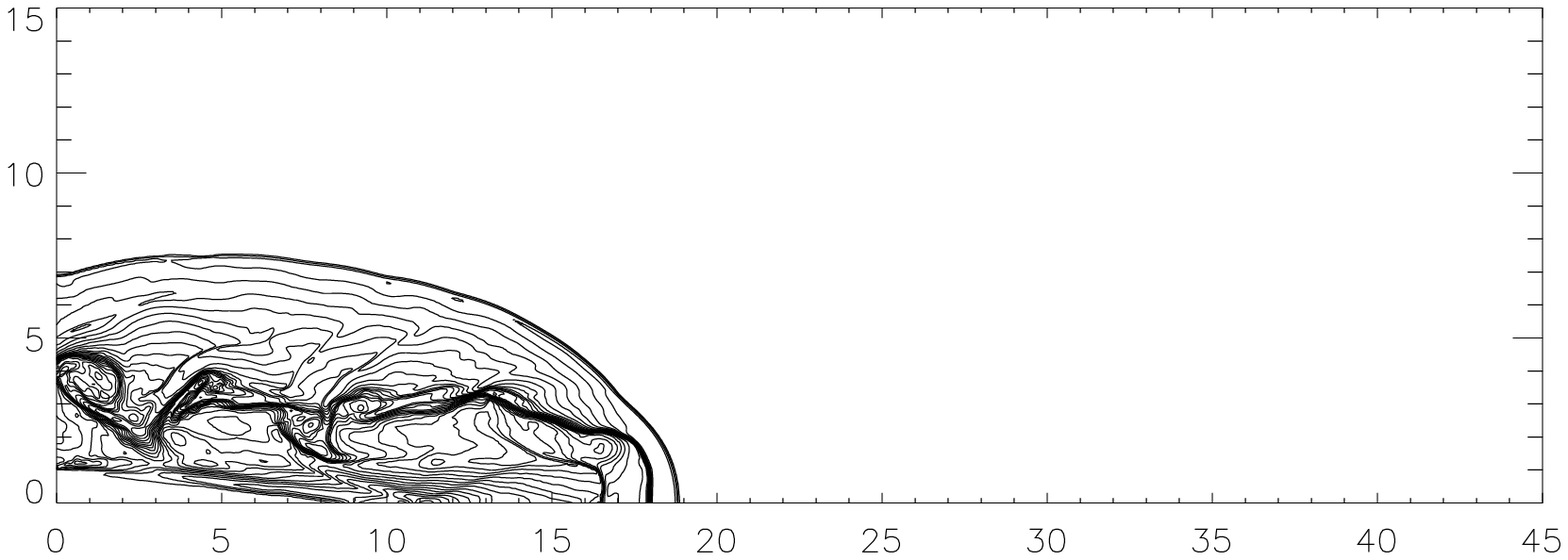}\\
\includegraphics[width=8cm]{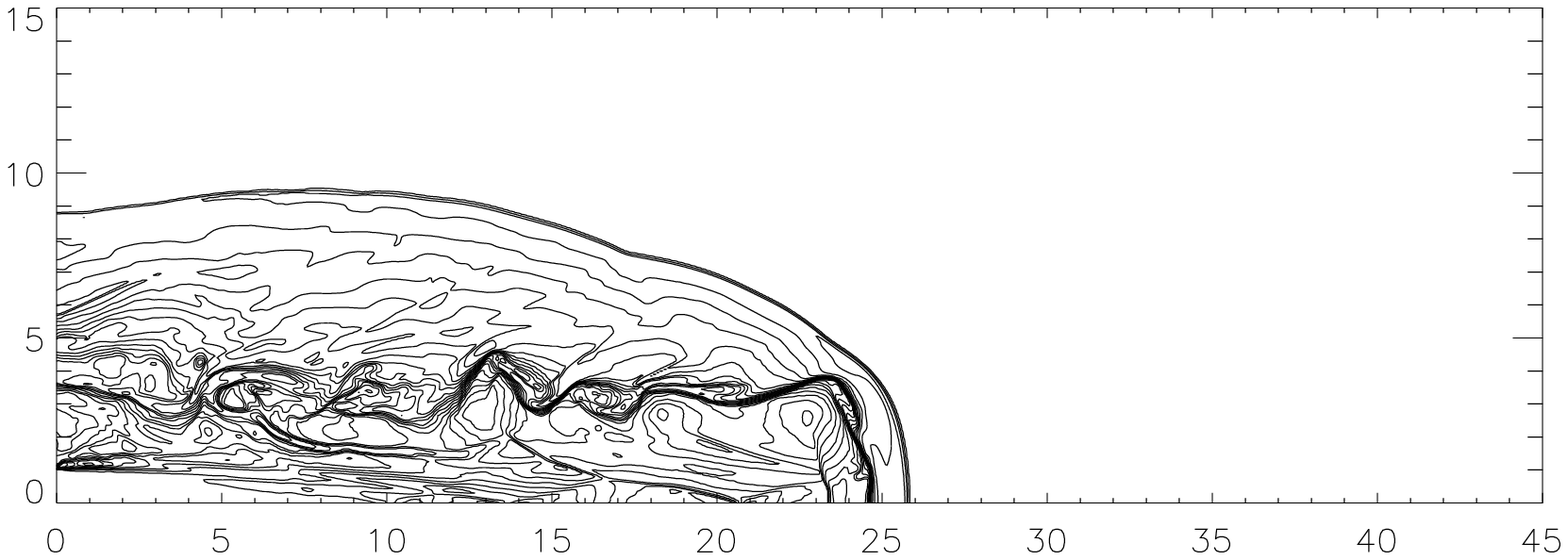}\\
\includegraphics[width=8cm]{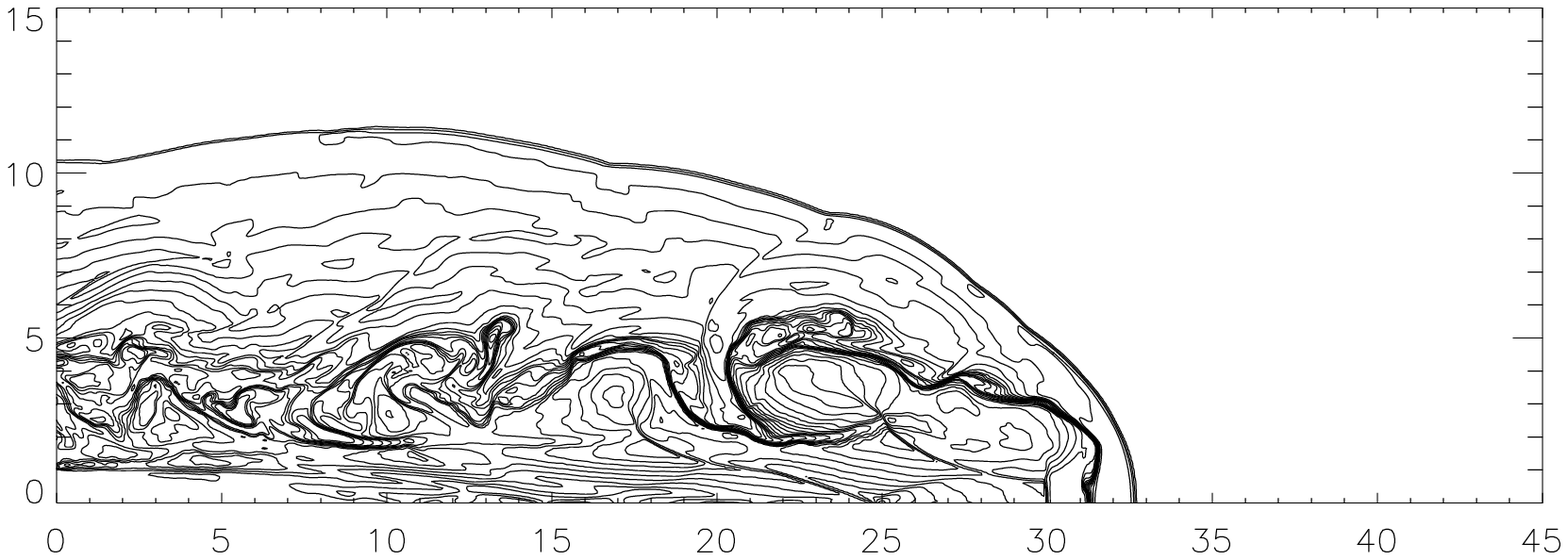}
\caption{Tadmor's scheme with PPM reconstruction in a relativistic jet
simulation. Snapshots of the rest mass density distribution (in
logarithmic scale) showing, from top to bottom, the propagation of
the jet at $t=20, 40, 60, 80$. In the last snapshot, the terminal
shock, the contact discontinuity and the bow shock are found at $z =
29.0, 31.2, 32.6$, respectively.}
\label{fig:jet}
\end{figure}

Figure~\ref{fig:jet} shows a series of snapshots of the rest mass density distribution 
covering the evolution of the jet up to $t=80$. This simulation is performed using 
cylindrical coordinates $(r,z)$ to discretize the computational domain, which is 45 
units long in the $z$-direction and 25 units wide in the $r$-direction. All structural 
features usually appearing in jet simulations are clearly identified in this figure. A 
supersonic beam extends from the jet nozzle to the point of impact with the ambient 
medium. At that point the jet ends up in a complex structure formed by a terminal
planar shock (Mach shock), a contact discontinuity separating the jet material from 
the shocked ambient medium, and a bow shock. This shock forms due to the supersonic 
propagation of the jet with respect to the ambient, allowing for the jet to evolve 
in a cavity of hotter and lighter shocked ambient material. Finally, the beam material 
which stops at the jet end forms a cocoon sourrounding the beam. The coccon is separated 
from the shocked ambient medium by a contact discontinuity in which Kelvin-Helmholtz 
instabilities develop. The velocity of advance of the jet in the ambient medium is 
governed by the balance of momentum at the jet/ambient impact region (see 
e.g.~\citet{marti97}). For the jet under consideration a theoretical advance speed of 
$0.44$ is expected (to be compared with a mean speed of 0.38 found in the simulation). 
The lateral expansion of the cavity as well as the evolution of pressure and density 
follows the theoretical model of~\citet{begelman89} with good accuracy.

\begin{figure}[t]
\centering
\includegraphics[width=8cm]{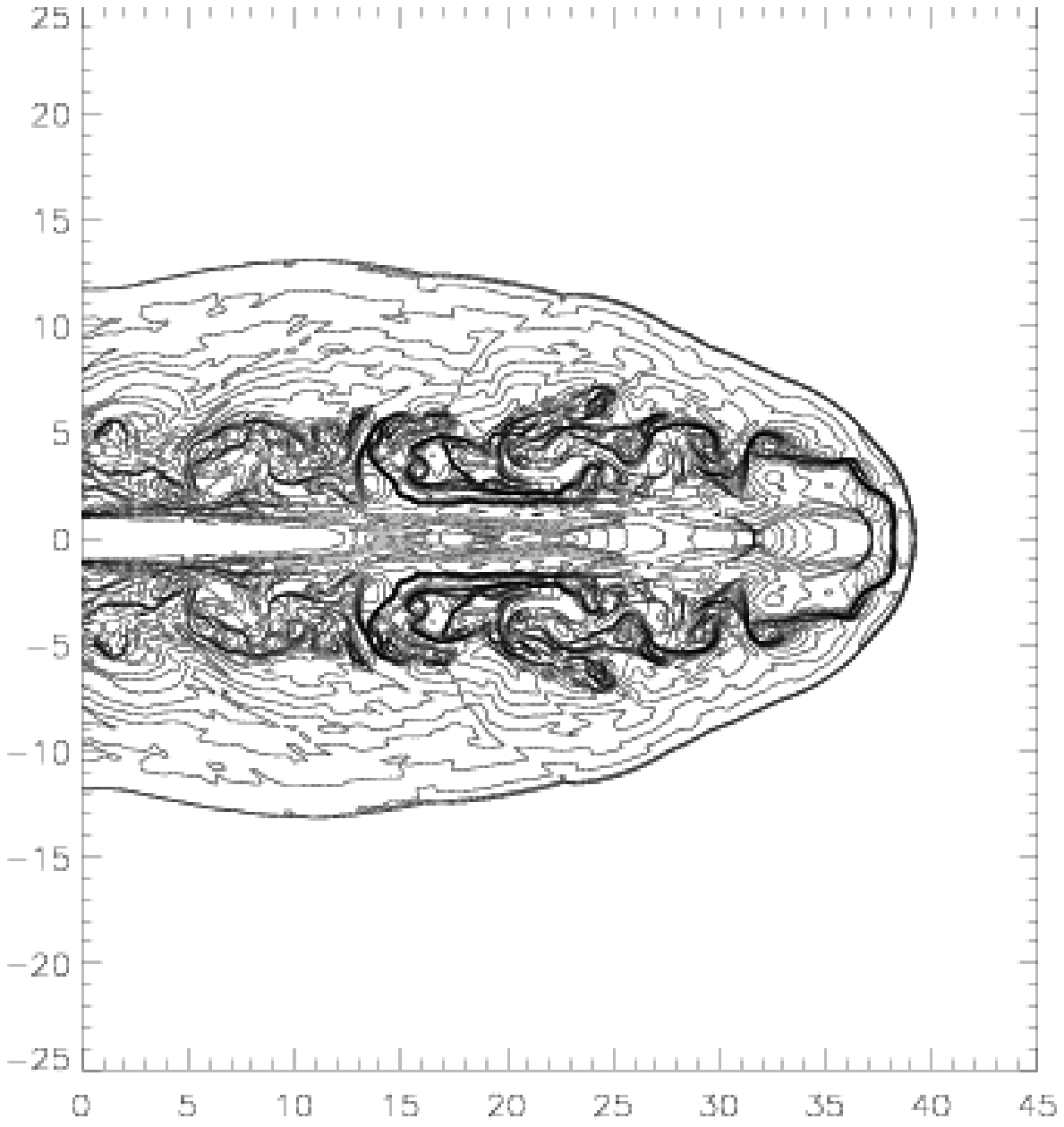}\\
\includegraphics[width=8cm]{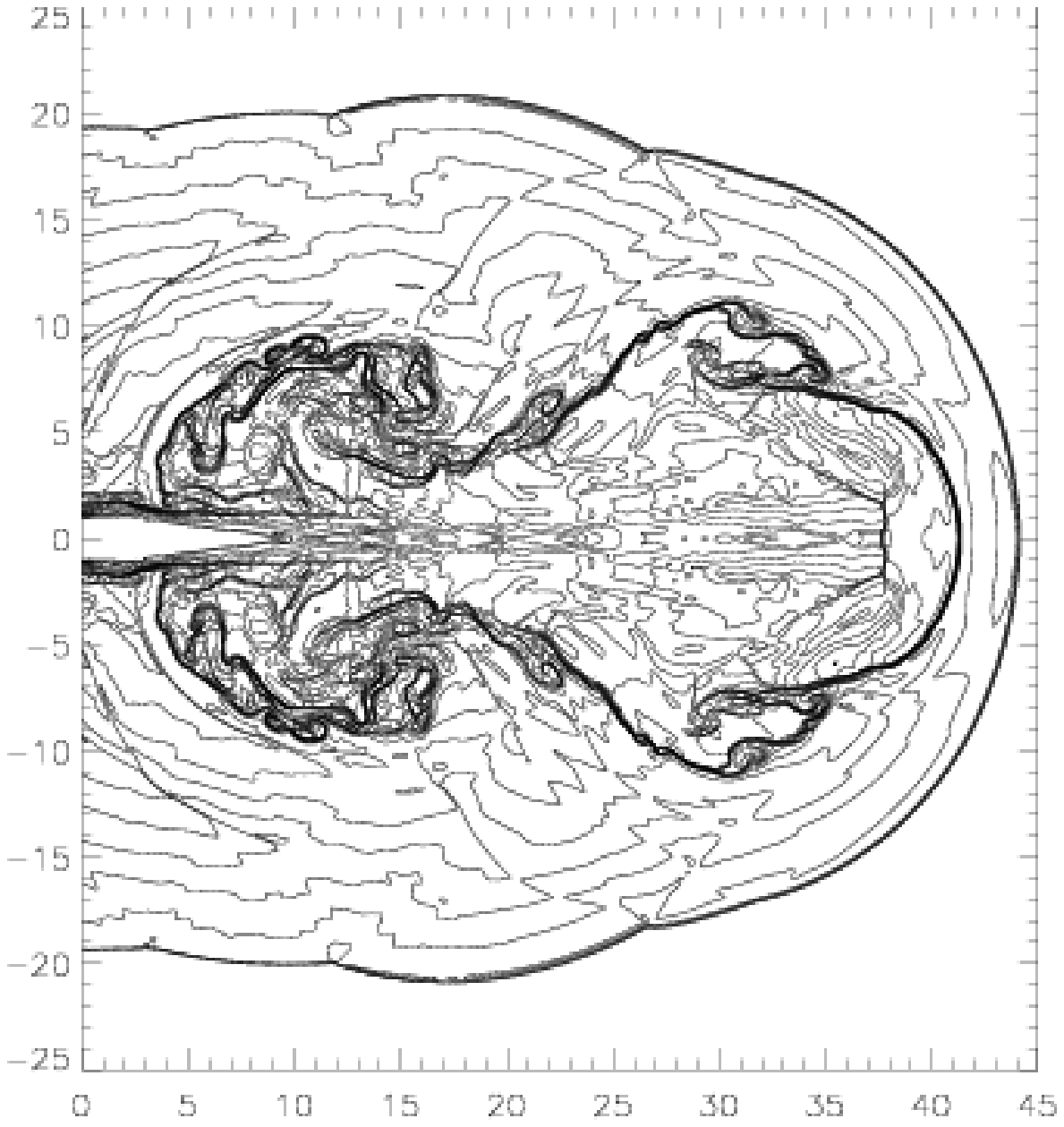}
\caption{Tadmor's scheme with PPM reconstruction in a relativistic jet
simulation. Rest mass density distribution (in logarithmic scale) at
$t=100$ for cylindrical (top panel) and planar (bottom panel) jets.}
\label{fig:jet_end}
\end{figure}

It is worth commenting on the absence of {\it carbuncle} ahead of the bow shock
in the simulations reported in Fig.~\ref{fig:jet}. The carbuncle is a numerical
pathology well known from blunt body simulations in supersonic gas dynamics, which 
manifests itself in the form of a small unphysical protuberance in front of the
bow shock. In this regard Tadmor's scheme is able to cleanly resolve (carbuncle-free)
the leading bow shock in the jet with an accuracy comparable to that of more 
sophisticated Godunov-type schemes such as Marquina, and clearly superior to that
of Roe-type approximate solvers, which sometimes admit this kind of spurious solution
as shown in~\citet{donat98}.

Figure~\ref{fig:jet_end} shows the last computed time of our model
($t = 100$) in 2D cylindrical and planar coordinates. The
morphological elements found in both cases are similar as well as the
dynamics governing the jet propagation. The change in geometry is the
responsible of the different aspect of the cavity (more elongated in
the case of the cylindrical jet).

%%%%%%%%%%%%%%%%%%%%%%%%%%%%%%%%%%%%%%%%%%%%%%%%%%%%%%%%%%%%%%%%%%%%%%%%%%%%%%%%%%%%%%%%%%%%
\section{Quantitative comparison}
\label{comparison}
%%%%%%%%%%%%%%%%%%%%%%%%%%%%%%%%%%%%%%%%%%%%%%%%%%%%%%%%%%%%%%%%%%%%%%%%%%%%%%%%%%%%%%%%%%%%

In this section we present a quantitative comparison between Tadmor's scheme and the HRSC 
methods based on (approximate) Riemann solvers we have used. To this aim we first compare 
the numerical and analytic solutions, and show the errors of the various methods for the
shock tube problem 3 and problem 6 (propagation of a blast wave with non-zero tangential 
velocities). We also show in each case the CPU time per numerical cell and iteration (TCI) 
in order to highlight the computational efficiency of our central scheme.

The results of the comparison are shown in Table~\ref{errors}. This table
reports the $L_1$ and $L_{\infty}$ norms for the density for Tadmor's scheme, 
HLLE, and Roe's approximate Riemann solvers. Results for both spatial reconstruction 
procedures (PPM and PHM) are also included. The $L_1$ norm errors are computed 
considering the entire computational domain, while the $L_{\infty}$ errors refer 
only to the part of the grid where the solution is smooth (i.e. the rarefaction 
wave). The largest errors occur in the postshock region. From this table we can 
quantify and confirm the quality of the results shown in the corresponding figures, 
i.e. that Tadmor's scheme has an accuracy comparable to that of Riemann solvers 
based HRSC methods.

\begin{center}
\begin{table}[t]
\centering
\caption{Errors in the shock tube problems 3 and 6 for HLLE and
Roe's approximate Riemann solvers, and Tadmor's central scheme using 
two different third order spatial reconstruction (CFL=0.5 and $N_x=400$).}
\begin{tabular}{cccccc}
\hline
Test & Scheme & \multicolumn{2}{c} {$L_1$ ($10^{-2}$)} & \multicolumn{2}{c} {$L_{\infty}$ ($10^{-3}$)} \\
 \hline
&  & PPM & PHM & PPM & PHM \\
 \hline
Problem 3 &  HLLE & 3.37 & 2.81 & 2.87 & 0.67 \\
&  Roe & 3.39 & 2.95 & 3.44 & 1.31 \\
&  Tadmor  & 3.41 & 3.41 & 2.58 & 1.04 \\
  \hline
Problem 6 &  HLLE & 22.0 & 21.7 & 3.53 & 1.20 \\
&  Roe & 21.7 & 21.6 & 2.63 & 1.20 \\
&  Tadmor  & 25.2 & 22.4 & 0.92 & 2.69 \\
  \hline
\end{tabular}
\label{errors}
\end{table}
\end{center}

Correspondingly, Table~\ref{TCI} displays the TCI for Tadmor's scheme and Marquina's 
flux formula, which allows us to check their computational efficiency. We note first 
that when writing the numerical code we did not take special care in optimization issues 
other than in its most quite apparent aspects. Therefore, the numbers displayed in 
Table~\ref{TCI} have to be taken as approximate numbers for a standard implementation of 
a hydrodynamics code, amenable to be improved under code optimization. From this table we 
can see that Tadmor's scheme, as expected, consumes less time in doing the calculations than 
a method based on the characteristic structure of the equations, roughly a factor of 2 for 
one-dimensional problems, a factor 4 for problem 6 where tangential flow velocities are also 
involved in the computation, and a factor 3 for a two-dimensional problem (namely, the 
flat-faced step test). 

\begin{center}
\begin{table}[t]
\centering
\caption{CPU time per numerical cell and iteration for Tadmor's
central scheme and Marquina's flux formula for the
different regimes considered in the text, using third order schemes for
the cell reconstruction and time update.} 
\begin{tabular}{cccc}
\hline
 \multicolumn{2}{c}{} & \multicolumn{2}{c}{TCI ($\mu$s)}\\
\hline
 Case & \# Zones & Marquina & Tadmor \\
 \hline
  1D & 400 x 1 x 1    & 17.0 & 8.0 \\
  1.5D  & 400 x 1 x 1 & 32.4 & 8.8 \\
  2D & 120 x 40 x 1   & 72.1 & 22.7 \\
  \hline
\end{tabular}
\label{TCI}
\end{table}
\end{center}

Some additional comments are in order: firstly, the TCI for Tadmor's scheme remains essentially 
unchanged when going from 1D to 1.5D (from 8.0 $\mu$s to 8.8 $\mu$s), which indicates the 
negligible impact of the computation of the numerical flux in the overall number of operations
in the code. Correspondingly, the factor of two difference in Marquina's flux formula (from 17.0 
$\mu$s to 32.4 $\mu$s) is consistent with the large relative weight of the numerical flux step 
in the overall computation in this algorithm (notice, in particular, that the numerical viscosity 
matrix changes its dimension from $3\times3$ to $5\times 5$). Secondly, when going from 1.5D to 
2D the TCI in Marquina's scheme increases by roughly a factor 2.2, while this factor is of the 
order of 2.5 in the case of Tadmor. A factor two can be easily understood due to the presence
of the additional dimension in the code, which implies an additional ``sweep" in the corresponding 
spatial dimension, the cell reconstruction and the numerical flux computation being also computed 
twice. This factor can be further modified by adding and subtracting two less important factors, 
namely the access to memory (which is slower as the arrays are larger) and the time update and 
recovery procedure which are done simultaneously for the two spatial dimensions.

Finally, we show in Table~\ref{table:convergence} the errors of the density under
grid refinement using the discrete $L_1$ norm. The results reported in this table
correspond to problem 4 at $t=0.4$, and they allow for a comparison of the various
schemes (Marquina, HLLE, and Tadmor) and reconstruction procedures (PPM and PHM). The
last two columns of the table indicate the convergence rate of the method. As the
grid is refined it can be seen how the convergence rate reaches an order of accuracy
of roughly one. This is the expected value for problems where discontinuities are
present. Furthermore, our result is also in very good agreement with the results of
\citet{mm96} where a relativistic extension of PPM was used in conjunction with the
exact Riemann solver (see their Table IV). 

\begin{center}
\begin{table}[t]
\centering
\caption{$L_1$ norm errors of the density and convergence rate ($r$) under grid
refinement for problem 4 at $t=0.4$. Results for three schemes (Marquina, HLLE,
and Tadmor) and two reconstruction procedures (PPM and PHM) are shown.}
\begin{tabular}{crcccc}
\hline
 Scheme & N & \multicolumn{2}{c}{$L_1 (10^{-2})$} & \multicolumn{2}{c}{r} \\
 \hline
  & & PPM & PHM & PPM & PHM \\
 \hline
  Marquina & 50 & 14.8 & 23.2 & | & | \\
           & 100 & 21.0 & 18.5 & -0.50 & 0.30 \\
           & 200 & 16.1 & 15.1 &  0.38 & 0.32 \\
           & 400 & 8.99 & 11.4 & 0.84 & 0.41 \\
           & 800 & 4.22 & 7.22 & 1.09 & 0.66 \\
           & 1600 & 2.22 & 3.87 & 0.93 & 0.90 \\
           & 3200 & 1.04 & 2.06 & 1.09 & 0.91 \\
 \hline
  HLLE & 50 & 14.8 & 23.3 & | & | \\
       & 100 & 20.9 & 17.0 & -0.50 & 0.29 \\
       & 200 & 16.1 & 15.1 & 0.38 & 0.33 \\
       & 400 & 8.99 & 11.4 & 0.84 & 0.41 \\
       & 800 & 4.22 & 7.22 & 1.09 & 0.66 \\
       & 1600 & 2.22 & 3.87 & 0.93 & 0.90 \\
       & 3200 & 1.04 & 2.06 & 1.09 & 0.91 \\
 \hline 
  Tadmor & 50 & 14.7 & 23.6 &  | & | \\
         & 100 & 20.7 & 19.1 & -0.49 & 0.30 \\
         & 200 & 16.0 & 15.2 & 0.37 & 0.33 \\
         & 400 & 8.94 & 11.0 & 0.84 & 0.47 \\
         & 800 & 4.19 & 7.26 & 1.09 & 0.60 \\
         & 1600 & 2.21 & 3.89 & 0.92 & 0.90 \\
         & 3200 & 1.04 & 2.07 & 1.09 & 0.91 \\
 \hline
\end{tabular}
\label{table:convergence}
\end{table}
\end{center}

\begin{figure}[t]
  \centering
  \includegraphics[angle=-90,width=8.5cm]{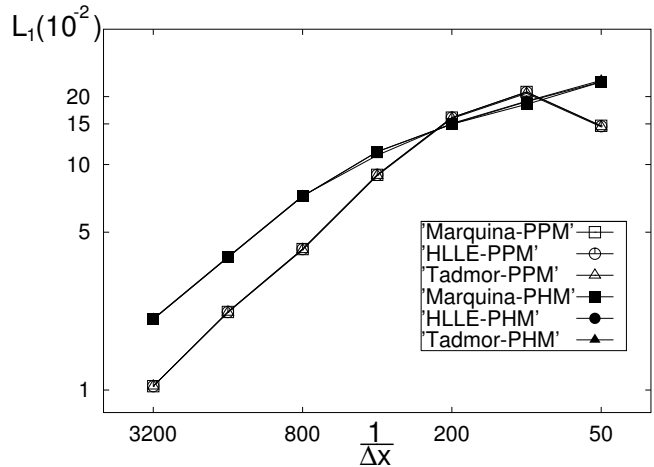}
  \caption{$L_1$ norm errors of the density under grid refinement for problem 4 at 
$t=0.4$. Open (respectively, filled) symbols correspond to results obtained with
PPM (respectively, PHM). Results for three schemes (Marquina (squares), HLLE (circles),
and Tadmor (triangles)) are shown. For a given grid resolution the accuracy of the 
solution shows no dependence on the numerical scheme and a strong dependence on the 
cell-reconstruction.}
  \label{Convergence}
\end{figure}

Two important conclusions can be drawn from Table~\ref{table:convergence}: firstly, the 
transition to the converged solution is faster with PPM than with PHM, especially for grid
resolutions finer than $1/400$. The corresponding $L_1$ norm errors are also systematically
smaller when using the PPM cell-reconstruction routines, roughly a factor of two better 
than PHM when the number of zones is $\ge 400$. Secondly, both the convergence rate and 
the errors are pretty much {\it independent} of the scheme (Marquina, HLLE, or Tadmor), 
and irrespective of the grid resolution employed. This result indicates that for a given 
scheme (either upwind or central) {\it written in conservation form}, it is the reconstruction 
what greatly helps in order to gain accuracy. These conclusions can also be inferred from 
Fig.~\ref{Convergence} which shows the convergence rate in logarithmic scale. It is however 
important to stress that this result applies to the specific test we have considered (problem 
4), for which the wave structure of the solution is particularly simple and the errors are 
dominated by the presence of discontinuities. It remains to be seen if our conclusion still 
holds in multidimensional problems involving more complex flows and wave interactions (not 
necessarily aligned with the computational grid) as those appearing, for instance, when 
simulating astrophysical jets.

%%%%%%%%%%%%%%%%%%%%%%%%%%%%%%%%%%%%%%%%%%%%%%%%%%%%%%%%%%%%%%%%%%%%%%%%%%%%%%%%%%%%%%%%%%%%%%%%%%%%%%%%%%
\section{Summary}
\label{section:summary}
%%%%%%%%%%%%%%%%%%%%%%%%%%%%%%%%%%%%%%%%%%%%%%%%%%%%%%%%%%%%%%%%%%%%%%%%%%%%%%%%%%%%%%%%%%%%%%%%%%%%%%%%%%

In this paper we have assessed the validity of a particular finite-difference central 
scheme in conservation form developed by \cite{kurganov}, for the solution of the
relativistic hydrodynamic equations. The computations have been restricted to one and 
two spatial dimensions in flat spacetime. Standard numerical experiments such as shock 
tubes, the shock reflection test, and the relativistic version of the flat-faced step 
test have been performed. As an astrophysical application of the central scheme we use 
we have presented two-dimensional simulations of the propagation of relativistic jets using 
both Cartesian and cylindrical coordinates. The simulations have shown the capabilities 
of Tadmor's scheme to yield satisfactory results, comparable to those obtained by HRSC 
schemes based on Riemann solvers, even well inside the ultrarelativistic regime. We have 
proposed to use high-order reconstruction procedures such as those provided by the PPM 
scheme \citep{colella} and the PHM scheme \citep{marquina}. This is essential to keep 
the inherent diffusion of central schemes at discontinuities at reasonably low levels.

The novelty of our approach (shared by recent earlier works in the literature 
\citep{delzanna,anninos}) relies on the absence of Riemann solvers in the solution 
procedure. Earlier pioneer approaches in the field of relativistic hydrodynamics 
\citep{norman,centrella} used standard finite-difference schemes in conjunction with
artificial viscosity terms to stabilize the solution across discontinuities. Those 
approaches, however, only succeeded in obtaining accurate results for moderate values 
of the Lorentz factor ($W\sim 2$). A key feature lacking in those earlier investigations 
was to write the system of equations and the numerical scheme in conservation form. To 
the light of our findings and, in addition, to the recent results reported by 
\citet{anninos} where artificial techniques have also been considered in tandem with 
central schemes, it appears that this was the ultimate reason preventing the extension 
of the computations to the ultrarelativistic regime.

The use of high-order central schemes for nonlinear hyperbolic systems of 
conservation laws is increasing in recent years since the seminal work in 
the second half of the 1980s \citep{davis,roe,yee1,tadmor}. Anile and 
coworkers \citep{anile} applied the central scheme SLIC in the context of the
time-dependent hydrodynamical semiconductor equations obtaining satisfactory 
results as well for a system whose characteristic information is not available. 
In the context of special and general relativistic MHD \citet{koide0,koide} applied 
a second-order central scheme with nonlinear dissipation developed by \citet{davis} 
to the study of relativistic extragalactic jets and black hole accretion. This scheme 
has been lately applied by \citet{mizuno} in general relativistic MHD simulations of 
the gravitational collapse of a magnetized rotating massive star as a model of gamma ray 
bursts. Also recently \citet{delzanna} and \citet{anninos} have assessed two different 
high-order central schemes in relativistic hydrodynamics, obtaining results as accurate 
as those of upwind HRSC schemes in standard tests.

The theoretical knowledge of the characteristic information of any hyperbolic 
system of equations is the key ingredient guiding the construction of HRSC upwind 
schemes. The different Riemann solvers available in the literature, either exact 
or approximate (see \citet{mm99} for an up-to-date list in relativistic 
hydrodynamics), all use in the solution procedure the eigenvalues (characteristic 
speeds) and/or the eigenvectors (characteristic fields) of the Jacobian matrices 
of the system. Such solvers provide a minute amount of diffusion across 
discontinuities while at the same time capture the jumps in a self-consistent way 
thanks to the implicit use of the Rankine-Hugoniot jump conditions (the so-called
shock-capturing property). Having such information available is also very important 
for the issue of imposing boundary conditions in regions where {\it a priori} there 
could exist some ambiguity. The knowledge of the behavior of the characteristic 
speeds and, therefore, the local directionality of the flow, greatly simplifies this 
task.

Needless to say, the alternative of using high-order central schemes as the one 
discussed here instead of upwind HRSC schemes becomes apparent when the spectral 
decomposition of the hyperbolic system under consideration is not known. The 
straightforwardness of a central scheme makes its use very appealing, especially in
multi-dimensions where computational efficiency is an issue. Perhaps the most 
important example in relativistic astrophysics is the system of (general) relativistic 
magnetohydrodynamic equations. Despite some recent progress has been accomplished in
recent years \citep{romero,balsara,komissarov}, much more work is certainly needed 
concerning their solution with HRSC schemes based upon Riemann solvers. Meanwhile, 
an obvious choice is the use of central schemes.

\begin{acknowledgements}
We thank Miguel-Angel Aloy for a careful reading of the manuscript.
This research has been supported by the Spanish Ministerio de
Ciencia y Tecnolog\'{\i}a (AYA2001-3490-C02-C01) and by the EU
Programme ``Improving the Human Research Potential and the
Socio-Economic Knowledge Base" (Research Training Network Contract
HPRN-CT-2000-00137). One of the authors (A.L.-S.) wants to dedicate 
this work to his father.
\end{acknowledgements}

\bibliographystyle{aa}
\bibliography{paper}

\end{document}